\pdfoutput=1 
\documentclass[pdfusetitle,aps,prl,twocolumn,superscriptaddress]{revtex4-2}

\usepackage{amssymb}
\usepackage{amsmath}
\usepackage[svgnames,dvipsnames,x11names]{xcolor}
\usepackage{hyperref}
\hypersetup{
colorlinks=true,
citecolor= NavyBlue,
linkcolor= Maroon,
urlcolor= NavyBlue,
pdfauthor={},
pdftitle={},
pdfsubject={}
}

\def \sI {{\sf I}}
\def \sJ {{\sf J}}
\def \sT {{\sf T}}
\def \cG {{\cal G}}

\def\wt{\widetilde}

\def \k {{\vec k}}

\def \hs {\hskip 1pt}

\def \cO {{\cal O}}

\def \LA {{\langle}}
\def \RA {{\rangle}}

\def \cL {{\cal L}}

\def \k {{\boldsymbol{k}}}
\def \n {{\boldsymbol{n}}}

\def \q {{\boldsymbol{q}}}

\def \x {{\boldsymbol{x}}}

\def \intS {\int_{S^2}}

\newcommand{\Gauss}[4]{ {}_2F_1\Bigg[\begin{array}{c} #1,\hs #2 \\[2pt] #3 \end{array}\Bigg|\, #4\Bigg]}

\newcommand{\HYPqR}[6]{ {}_3\wt F_2\Bigg[\begin{array}{c} #1,\hs #2,\hs #3 \\[2pt] #4,\hs #5 \end{array}\Bigg|\, #6\Bigg]}

\newcommand{\HYPwR}[8]{ {}_4\wt F_3\Bigg[\begin{array}{c} #1,\hs #2,\hs #3 ,\hs #4 \\[2pt] #5,\hs #6,\hs #7  \end{array}\Bigg|\, #8\Bigg]}

\def\nn{\nonumber\\}

\begin{document}

\title{Tensor Integrals in the Large-Scale Structure}

\author{Hayden Lee}
\affiliation{Kavli Institute for Cosmological Physics, University of Chicago, Chicago, IL 60637, USA}
\affiliation{Department of Physics and Astronomy, University of Pennsylvania, Philadelphia, PA 19104, USA}

\begin{abstract}
We present a new method for evaluating tensor integrals in the large-scale structure.
Decomposing a $\Lambda$CDM-like universe into a finite sum of scaling universes using the FFTLog, we can recast loop integrals for biased tracers in the large-scale structure as certain tensor integrals in quantum field theory.
While rotational symmetry is spontaneously broken by the fixed reference frame in which biased tracers are observed, the tensor structures can still be organized to respect the underlying symmetry.
Projecting the loop integrands for scaling universes onto spherical harmonics, the problem effectively reduces to the evaluation of one-dimensional radial integrals, which can be solved analytically.
Using this method, we derive analytic expressions for the one-loop power spectrum, bispectrum, and trispectrum for arbitrary multipole moments in the basis of scaling universes.
\end{abstract}

\maketitle

\section{Introduction}

The study of the large-scale structure (LSS) of the universe is set to mark a new era in precision cosmology.
Upcoming galaxy surveys, such as DESI~\cite{DESI:2016fyo}, Euclid~\cite{EUCLID:2011zbd}, SPHEREx~\cite{SPHEREx:2014bgr}, LSST~\cite{LSST:2008ijt}, the Roman Space Telescope~\cite{Akeson:2019biv}, as well as the proposed MegaMapper~\cite{Schlegel:2022vrv}, will offer an unprecedentedly accurate view of the matter distribution in the universe.
These advances hold the potential to critically test and refine our understanding of dark matter, dark energy, and the primordial universe.

To fully exploit this wealth of data, precise and accurate modeling of gravitational clustering in the quasi-linear regime is crucial. 
A systematic framework for this study is provided by the effective field theory (EFT) of the LSS~\cite{Baumann:2010tm, Carrasco:2012cv,Cabass:2022avo, Ivanov:2022mrd}. 
Within this framework, computing subleading corrections to linear theory amounts to evaluating certain momentum loop integrals.
Over the past decade, substantial progress has been made in advancing the EFT approach for computing loop corrections to the power spectrum~\cite{Carrasco:2013sva, Carrasco:2013mua,Pajer:2013jj,Senatore:2014via,Baldauf:2015aha,Konstandin:2019bay}, bispectrum~\cite{Angulo:2014tfa,Baldauf:2014qfa,Eggemeier:2018qae,DAmico:2022ukl}, and trispectrum~\cite{Bertolini:2015fya, Bertolini:2016bmt, Steele:2021lnz}.

Despite being conceptually straightforward, the computational demands of accurate calculations are challenging for LSS loop integrals.
Notably, every loop integrand depends on the initial matter power spectrum that is only known in numerical form, which precludes direct analytic integration.
Enhancing the computation efficiency of these loop integrals is crucial, as observables need to be accurately computed millions of times across different cosmologies to effectively scan the wide parameter space from Monte Carlo Markov Chain (MCMC) analyses. 
Unfortunately, brute-force numerical evaluation of these integrals is inadequate for this purpose.

This need has spurred the development of fast computational methods for LSS loop integrals. 
One notable approach is the FFTLog~\cite{Simonovic:2017mhp} (see~\cite{Anastasiou:2022udy} for a related approach and earlier works~\cite{Schmittfull:2016jsw,McEwen:2016fjn, Schmittfull:2016yqx,Fang:2016wcf}), along with its application to full-sky angular space~\cite{Assassi:2017lea,  GrasshornGebhardt:2017tbv, Schoneberg:2018fis, Lee:2020ebj, Chen:2021vba}.
The key idea behind this method is to decompose the numerical integrand into a sum of simpler analytic functions, with individual terms resembling standard loop integrals in quantum field theory (QFT). 
Consequently, many of these integrals become analytically solvable, dramatically enhancing computational speed. 
A major advantage of this approach over traditional numerical methods is that the integrals are computed independently of cosmological parameters, allowing them to be recycled for different cosmologies.
Thanks to these new techniques, EFT analyses of the BOSS data have been successfully performed in~\cite{Ivanov:2019pdj, Colas:2019ret}.

This progress will further advance with future observations that probe higher redshifts, including other biased tracers of the underlying matter density, such as the Lyman-alpha forest and the 21-cm transition of neutral hydrogen. 
These tracers can also be treated within the same EFT framework~\cite{Chen:2021rnb,Qin:2022xho,Ivanov:2023yla}.
Since these biased tracers are naturally observed in redshift space, they are sensitive to redshift-space distortions and line-of-sight (LOS) dependent selection effects.
As a result, loop integrals for LSS tracers are intrinsically ``tensor integrals,'' with the tensors capturing the angular dependence of momenta relative to the LOS direction. 
This introduces additional complexities compared to the ``scalar integrals'' for the underlying density fields evaluated in real space.

Similar tensor integrals have been extensively studied in the context of scattering amplitudes~\cite{Henn:2014yza,Weinzierl:2022eaz}.
The well-known technique due to Passarino and Veltman~\cite{Passarino:1978jh} reduces tensor integrals to linear combinations of scalar integrals multiplied by independent tensor structures.
Although this method has been used in previous studies for redshift-space loop integrals~\cite{Chudaykin:2020aoj,Philcox:2022frc,DAmico:2022ukl}, this is not always ideal.
For instance, this approach involves matrix inversions and often leads to a proliferation of scalar integrals as the rank of the tensor increases, with individual integrals potentially having spurious divergences.
Moreover, the efficiency of tensor reduction is highly dependent on the choice of basis.
For amplitudes, despite a number of earlier studies~\cite{Davydychev:1991va,Tarasov:1996br,Bern:1993kr,Binoth:1999sp}, the efficient computation of tensor integrals remains an area of active exploration~\cite{Chen:2019wyb,Peraro:2020sfm,Lyubovitskij:2021ges, Anastasiou:2023koq,Goode:2024mci}. 
It is desirable to make comparable progress in the study of LSS observables.

In this {\it letter}, we address the challenge of evaluating tensor integrals in the LSS by organizing the tensor structures based on symmetry.
While rotational symmetry is spontaneously broken by the LOS direction, independent tensor structures can still be organized to respect the underlying symmetry.
This is achieved by projecting the integrands onto a basis of spherical harmonics, which is particularly simple in three dimensions. 
This effectively reduces the problem to one-dimensional radial integration, allowing for analytic treatment.
Concretely, we provide analytic expressions for certain tensor ``master integrals'' associated with the power spectrum, bispectrum, and trispectrum at one loop.
As we will demonstrate, a key benefit of our approach is its uniform applicability for all multipole moments, unlike previous methods that required computing each moment separately.

\section{Loops\! in\! the\! Large-Scale Structure}

In this section, we introduce the basic ingredients of our approach. 
We begin with a brief review of the perturbation theory of biased tracers in redshift space and the FFTLog decomposition of the power spectrum into a sum over scaling universes. 
We then describe the spherical harmonic projection of the loop integrands. 

\subsection{Biased Tracers in the Large-Scale Structure}

In the LSS, we typically observe the density $\delta_{\rm tr}$ of a biased tracer of the underlying matter density $\delta$.
In the quasi-linear regime, their relationship can be treated perturbatively through a bias expansion~\cite{Chan:2012jj,Assassi:2014fva,Senatore:2014eva,Desjacques:2016bnm}
\begin{align}
	\delta_{\rm tr}(\x,z) &= \sum_\cO b_\cO(z) \cO(\x,z)+\cdots\,,
\end{align}
where $\x$ is the spatial coordinate, $z$ is the redshift, $\cO$ denote operators that affect the formation and evolution of the tracer, $b_\cO$ are their associated bias parameters, and the ellipsis denotes stochastic contributions.
The operators $\cO$ are constructed from $\delta$, which can be solved perturbatively in Fourier space~\cite{Bernardeau:2001qr}.

The coordinate transformation from real space to redshift space introduces a dependence on the LOS direction $\hat\n$.
This can be systematically incorporated into the EFT, resulting in the following perturbative solution for the tracer field in redshift space~\footnote{Assuming the Einstein-de Sitter approximation and neglecting scale-dependent effects due to massive neutrinos, the time and momentum dependencies factorize. The scale dependence can be incorporated using, e.g.,~the polynomial approximation of the growth factor~\cite{Levi:2016tlf,Chen:2021vba}.}:
\begin{align} 
	\delta_{\rm tr}^{(\hat\n)}(\k,z)  &= \sum_{n=1}^\infty \frac{D^n(z)}{D^n(z_0)}\int_{\q_1}\cdots \int_{\q_n}(2\pi)^3\delta_{\rm D}(\textstyle{\k-\sum_{i=1}^n}\q_i)\nn
	&\quad\times Z_n(\q_1,\cdots,\q_n,\hat\n)\delta_0(\q_1)\cdots\delta_0(\q_n)\,,
\end{align}
where $\int_\q \equiv \int \frac{d^3q}{(2\pi)^3}$ denotes a three-dimensional momentum integral, $D(z)$ is the linear growth factor, $\delta_{\rm D}$ is the Dirac delta function, $Z_n$ are the redshift-space kernels, and $\delta_0$ is the linear matter density at $z=z_0$.
For an explicit form of $Z_n$ for galaxies, see~\cite{Perko:2016puo, Philcox:2022frc,DAmico:2022ukl}. 
For our purposes, it suffices to know the general functional form of $Z_n$ and their building blocks; these are
\begin{align}
	Z_n(\q_1,\cdots\q_n,\hat\n)\ \supset\ |\textstyle\sum_i \sigma_i \q_i|^{\pm 2}\,,\,\hat\q_i\cdot\hat\n\,,
\end{align}
where $\sigma_i\in\{-1,0,1\}$. 
In other words, the kernels are rational functions of the squared momentum magnitudes and polynomials of the angles $\hat\q_i\cdot\hat\n$.
For instance, the one-loop power spectrum includes contributions from integrals of the form
\begin{align}
	\int_{\q}\frac{(\hat\q\cdot\hat\n)^\ell}{q^{2n_1}|\k-\q|^{2n_2}} P_{\rm lin}(q)P_{\rm lin}(|\k-\q|)\,,
\end{align}
for $\ell=0,1,2,\cdots$, where $n_i\in\mathbb{Z}$, and $P_{\rm lin}$ is the linear power spectrum that arises from Wick contraction, $\LA \delta_0(\k)\delta_0(\k')\RA=P_{\rm lin}(k)(2\pi)^3 \delta_{\rm D}(\k+\k')$, with $k\equiv |\k|$.
Here, $P_{\rm lin}$ is a numerical function obtainable from Boltzmann codes, which forbids direct analytic computation of these integrals. 
Since brute-force numerical integration is not feasible in modern MCMC analyses, it motivates us to pursue an alternative approach.

\subsection{FFTLog Decomposition}

To accelerate the evaluation of LSS loop integrals, a new strategy was introduced in~\cite{Simonovic:2017mhp}, building on the FFTLog decomposition of the power spectrum~\cite{Hamilton:1999uv, McEwen:2016fjn,Schmittfull:2016jsw}. 
The central idea is to approximate the linear power spectrum in terms of complex power-law functions (scaling universes) as
\begin{align}
	P_{\rm lin}(k)  &\approx \sum_\nu c_\nu k^{\nu} \, ,
\end{align}
with some coefficients $c_\nu$, where $\nu\in\mathbb{C}$. 
Typically, $O(100)$ terms are sufficient to accurately capture the full shape of the linear power spectrum, including the baryon acoustic oscillations. 
Since $c_\nu$ are just numbers, the cosmology-dependent coefficients completely factorize outside the integrals, and the integrals need to be computed only once for a pre-selected set of $\nu$ parameters.
Evaluation of standard loop diagrams then boils down to simple matrix multiplication.

In redshift space, the perturbation theory kernels depend not only on momentum magnitudes but also on the angle $\hat\q\cdot\hat\n$, which makes evaluation of the integrals technically more involved. 
Expanding the redshift-space kernels, all integrals can be reduced to basic tensor integrals of the form~\footnote{It is sufficient to consider numerators of the form $(\hat\q\cdot\hat\n)^\ell$, since contraction with the external momenta can always be decomposed into pure scalar integrals using $\hat\q\cdot\k_i = \frac{1}{2q}(q^2+k_i^2-|\q - \k_i|^2)$.}
\begin{align}
	i_\ell^{\nu_1\nu_2}(\k,\hat\n)&\equiv \int_{\q}\frac{(\hat\q\cdot\hat\n)^\ell}{q^{2\nu_1}|\k-\q|^{2\nu_2}}\,,\label{iint}\\ 
	j_\ell^{\nu_1\nu_2\nu_3}(\k_1,\k_2,\hat\n)&\equiv \int_{\q} \frac{(\hat\q\cdot\hat\n)^\ell}{q^{2\nu_1}|\k_1-\q|^{2\nu_2}|\k_2+\q|^{2\nu_3}}\,,\label{jint}
\end{align}
for the one-loop power spectrum and bispectrum, respectively~\footnote{For standard galaxy biasing, multipoles up to $\ell=4$ and $\ell=6$ contribute to the one-loop power spectrum and bispectrum, respectively. 
However, additional selection effects, such as those for certain samples of galaxies~\cite{Desjacques:2018pfv, Tomlinson:2020xbf} and other tracers like the Lyman-alpha forest~\cite{Ivanov:2023yla}, necessitate including higher multipoles (up to $\ell=8$ for the one-loop power spectrum).}. 
These take the form of massless tensor integrals in QFT, with the ``propagators'' having complex exponents. 
For the monopole ($\ell=0$), analytic expressions for these integrals were derived in~\cite{Simonovic:2017mhp}.

For $\ell>0$, the standard technique for handling tensor integrals is to decompose them into independent tensor structures multiplied by scalar integrals~\cite{Chudaykin:2020aoj, Philcox:2022frc, DAmico:2022ukl, Ivanov:2023yla}. 
However, a naive such decomposition can result in a proliferation of scalar integrals and involve matrix inversions that can lead to numerical instabilities in certain kinematic regions.
Below, we show that there is a preferred choice of basis of tensor structures that circumvents these issues and greatly simplifies the resulting expressions.

\subsection{Spherical Harmonic Projection} 

Our main idea involves choosing a basis of tensor structures that respects the underlying rotational symmetry, as these tensors must be covariant under SO(3) rotations before contracting with $\hat\n$.
This choice allows us to easily evaluate the angular integrals, effectively reducing the problem to one-dimensional radial integration.
To accomplish this, we project all $\hat\q$-dependent factors of the loop integrand onto spherical harmonics using the {\it Funk-Hecke formula}~\cite{Funk1916, Hecke1917}. 
Since this formula is not widely used in cosmology and particle physics, we first outline a general formulation in $d$ dimensions before specializing to $d=3$. 

Spherical harmonics form an orthogonal basis for the spin-$\ell$ representation $\rho_{d}^{(\ell)}$ of the rotation group SO($d$). 
Using the addition theorem of spherical harmonics, any scalar function $f$ of the angle between two unit vectors $\hat\xi,\hat\eta\in S^{d-1}$ can be expanded as
\begin{align}
	f(\hat\xi\cdot\hat\eta)
	&= \sum_{\ell=0}^\infty \frac{| \rho_{d}^{(\ell)}| }{|S^{d-1}|}\lambda_\ell \,C_\ell^{(\frac{d-3}{2})}(\hat\xi\cdot\hat\eta)\,,\label{proj}
\end{align}
where 
$|\rho_{d}^{(\ell)}|=\binom{\ell+d-1}{\ell}-\binom{\ell+d-3}{\ell-2}$ 
is the dimension of $\rho_{d}^{(\ell)}$, $|S^{d-1}|=\frac{2\pi^{d/2}}{\Gamma(d/2)}$ is the surface area of the unit $(d-1)$-sphere, $C_\ell^{(\alpha)}$ is the Gegenbauer polynomial, and the coefficients $\lambda_\ell$ are evaluated by the projection integral
\begin{align}
	\lambda_\ell &= 
	\frac{|S^{d-2}|}{C_\ell^{(\frac{d-3}{2})}(1)} \int_{-1}^1dt \,(1-t^2)^{\frac{d-3}{2}} f(t)C_\ell^{(\frac{d-3}{2})}(t)\, .\label{FH}
\end{align}
For further mathematical background, see, e.g.,~\cite{atkinson2012spherical}.

The above expressions become particularly simple in ${d=3}$. 
For LSS applications, we will apply this formula to the integrands in \eqref{iint} and \eqref{jint}. 
Projecting the angle-dependent factor onto the Legendre basis gives
\begin{align}
	\frac{1}{|\k - \q|^{2\nu}} &= \sum_{\ell=0}^\infty  \frac{2\ell+1}{4\pi}\lambda_\ell^\nu(k,q) 
\cL_\ell(\hat \k\cdot\hat\q)\,,
\label{powerlawproj}
\end{align}
where $\cL_\ell$ is the Legendre polynomial of degree $\ell$ and the $k,q$-dependent coefficients are given by
\begin{equation}
	\lambda_\ell^\nu(k,q) = 2\pi  \int_{-1}^1 dt\,  (k^2+q^2-2kqt)^{-\nu}\cL_\ell(t) \,.
\end{equation} 
Using the Rodrigues formula $\cL_\ell(t) = \frac{\pi}{2^{\ell-1}\ell!}\frac{d^\ell}{dt^\ell}(t^2-1)^\ell$ and integrating by parts $\ell$ times, this integral evaluates to
\begin{align}
	\lambda_\ell^\nu(k,q)
	&= \frac{2\pi^{\frac{3}{2}}\Gamma(1-\nu)}{\Gamma(\ell+\frac{3}{2})\Gamma(1-\nu-\ell)}\frac{(-kq)^{\ell}}{(k^2+q^2)^{\ell+\nu}}\nn
	&\qquad\times \Gauss{\tfrac{\ell+\nu}{2}}{\tfrac{\ell+\nu+1}{2}}{\ell+\tfrac{3}{2}}{\frac{4k^2q^2}{(k^2+q^2)^2}}\,,\label{lam}
\end{align}
where ${}_2F_1$ is the Gauss hypergeometric function. 
The argument of ${}_2F_1$ lies within $[0,1]$ and the series expansion around the origin always converges for $k\ne q$. 
For $k= q$, the hypergeometric series converges for $\Re[\nu]<1$.
The above hypergeometric function with the specified parameters is non-generic and satisfies interesting identities, see Appendix~\ref{app:deriv}.

\section{One-Loop Power Spectrum}

To exploit the rotational symmetry inherent in the problem, we consider the following master integral for the one-loop power spectrum:
\begin{align}
	I_\ell^{\nu_1\nu_2}(\k,\hat\n) \equiv  \int_\q  \frac{\cL_\ell(\hat\q\cdot\hat\n)}{q^{2\nu_1}|\k-\q|^{2\nu_2}}\,,\label{Iint}
\end{align}
with the Legendre polynomial in the numerator instead of a monomial as in \eqref{iint}. 
Although the two choices are trivially related, starting with the Legendre polynomial allows us to easily perform the angular integration.
After projecting $|\k-\q|^{-2\nu_2}$ onto spherical harmonics using \eqref{powerlawproj}, the angular integration can be trivially done, giving $\cL_\ell(\hat\k\cdot\hat\n)$. This then leaves us with the radial integral
\begin{equation}
	I_\ell^{\nu_1\nu_2}(\k,\hat\n) =\cL_\ell(\hat\k\cdot\hat\n)\int_0^\infty \frac{q^2dq}{(2\pi)^3} \,q^{-2\nu_1}\lambda_{\ell}^{\nu_2}(k,q)\,.
\end{equation}
The remaining $q$ integral can be done straightforwardly (see Appendix~\ref{app:deriv}), and we obtain
\begin{align} 
&I_\ell^{\nu_1\nu_2}(\k,\hat\n)= \cL_\ell(\hat\k\cdot\hat\n)k^{3-2\nu_{12}}{\sf I}_\ell^{\nu_1\nu_2}\,,\\[3pt]
	&{\sf I}_\ell^{\nu_1\nu_2}= \frac{1}{8\pi^{\frac{3}{2}}}\frac{\Gamma(\frac{3+\ell}{2}-\nu_1)\Gamma(\frac{3}{2}-\nu_2)\Gamma(\nu_{12}-\frac{3-\ell}{2})}{\Gamma(\nu_1+\frac{\ell}{2})\Gamma(\nu_2)\Gamma(3+\frac{\ell}{2}-\nu_{12})}\,,\label{sI} 
\end{align}
where $\nu_{1\cdots n}\equiv \nu_{1}+\cdots+\nu_{n}$. 
This generalizes the monopole formula for $\ell=0$ given in~\cite{Scoccimarro:1996jy, Pajer:2013jj, Simonovic:2017mhp} to arbitrary $\ell$.

We can easily rewrite the result back in terms of the monomial basis using
\begin{align}
	x^\ell =\sum_{p=0}^{\lfloor \ell/2\rfloor}\frac{\ell!(2\ell-4p+1)}{2^p p!(2\ell-2p+1)!!} \cL_{\ell-2p}(x)\,.\label{monomial} 
\end{align} 
For concreteness, let us write out the explicit relations (for even $\ell$) between the integrals \eqref{iint} and \eqref{Iint}:
\begin{align}
	i_2  &= \frac{2}{3}I_2+\frac{1}{3}I_0\,,\\
	i_4  &= \frac{8}{35}I_4+\frac{4}{7}I_2+\frac{1}{5}I_0\,,\\
	i_6  &= \frac{16}{231}I_6+\frac{24}{77}I_4+\frac{10}{21}I_2+\frac{1}{7}I_0\,,\\
	i_8  &= \frac{128}{6435}I_8+\frac{64}{495}I_6+\frac{48}{143}I_4+\frac{40}{99}I_2+\frac{1}{9}I_0\,,
\end{align}
where we have suppressed the arguments and superscripts for brevity. 
One can verify that these are equivalent to the results shown in~\cite{Chudaykin:2020aoj, Ivanov:2023yla, Linde:2024uzr}, which involve many scalar integrals for higher multipoles (e.g., $O(100)$ terms for $\ell=8$)~\footnote{To match these results requires using recursive relations between gamma functions. As a simple check, our ${\sf I}_{\ell=1}^{\nu_1\nu_2}$ is related to $A_1=\frac{1}{2}({\sf I}_0^{\nu_1\nu_2}+{\sf I}_0^{\nu_1-1,\nu_2}-{\sf I}_0^{\nu_1,\nu_2-1})$ defined in \cite{Chudaykin:2020aoj} by ${\sf I}_{1}^{\nu_1-\frac{1}{2},\nu_2}=A_1$.}.
In contrast, our new formula expresses $i_\ell^{\nu_1\nu_2}$ as a sum of a few scalar integrals. 

\section{One-Loop Bispectrum}

The one-loop bispectrum in redshift space was studied in~\cite{Chudaykin:2020aoj, Philcox:2022frc, DAmico:2022ukl, Ivanov:2023yla} using the standard tensor reduction method.
Here, we will consider the following master integral in the Legendre basis:
\begin{equation}
	J_\ell^{\nu_1\nu_2\nu_3}(\k_1,\k_2,\hat\n)\equiv\!\int_\q \frac{\cL_\ell(\hat\q\cdot\hat \n)}{q^{2\nu_1}|\k_1-\q|^{2\nu_2}|\k_2+\q|^{2\nu_3}}\label{Jint}\,.
\end{equation}
Again, we can use \eqref{monomial} to easily go back to the monomial basis in \eqref{jint}.
To proceed, we first project both ${|\k_1-\q|^{-2\nu_2}}$ and $|\k_2+\q|^{-2\nu_3}$ onto spherical harmonics, and then perform the angular integration.
After factoring out an overall dimensionful factor $k_1^{3-2\nu_{123}}$, this gives
\begin{align}
&J_\ell^{\nu_1\nu_2\nu_3}(\hat\k_1,\hat\k_2,\hat\n) \nn
&\quad = k_1^{3-2\nu_{123}}\!\!\sum_{L_2,L_3=0}^\infty S^{L_2L_3\ell}(\hat\k_1,\hat\k_2,\hat\n) \sJ_{L_2L_3}^{\nu_1\nu_2\nu_3}(x)\,,\label{Jint2}
\end{align}
where $x\equiv k_2/k_1$ and the angular dependence is captured by the {\it tripolar scalar harmonics}~\cite{VMK,Szapudi:2004gh}
\begin{align}
	&  S^{L_2L_3\ell}(\hat\k_1,\hat\k_2,\hat\n)\equiv \frac{4\pi}{2\ell+1}\sum_{M_2=-L_2}^{L_2}\sum_{M_3=-L_3}^{L_3} \sum_{m=-\ell}^{\ell} \label{TSH}\\
	&\qquad\qquad\qquad\cG^{L_2L_3\ell}_{M_2M_3m}Y_{L_2M_2}^*(\hat\k_1)Y_{L_3M_3}^*(\hat\k_2)Y_{\ell m}^*(\hat\n)\,,\nonumber
\end{align}
which is a fully symmetric, orthogonal function of the angles between three unit vectors~\footnote{This can also be written manifestly as a function of the three angles by expressing it as a sum over a product of three Legendre polynomials of these angles. However, it is computationally more efficient to fix a coordinate system and evaluate~\eqref{TSH}.}, with the Gaunt integral $\cG^{L_2L_3\ell}_{M_2M_3m}$ defined in \eqref{gaunt} and $Y_{\ell m}$ is the spherical harmonics on $S^2$~\footnote{We adopt a slightly different normalization compared to~\cite{VMK}. The tripolar scalar harmonic \eqref{TSH} consists of 3$j$ symbols and can be computed very efficiently using numerical algorithms such as the one presented in~\cite{Johansson:2015cca}.}. 
The radial integral is given by
\begin{align}
	&\sJ_{L_2L_3}^{\nu_1\nu_2\nu_3}(x) \\
	&=(-1)^{L_3}\!\int_0^\infty \frac{q^2dq}{(2\pi)^3} q^{-2\nu_1} \lambda_{L_2}(\nu_{2};1,q)\lambda_{L_3}(\nu_{3},x,q)\,.\nonumber
\end{align}
This integral can be computed analytically in terms of a generalized hypergeometric function, whose detailed derivation is provided in Appendix~\ref{app:deriv}. The final result is given by
\begin{widetext}
\begin{align}
	\sJ_{L_2L_3}^{\nu_1\nu_2\nu_3}(x) &=  (-1)^{L_2}\Big(x^{3-2\nu_{13}} \,{\sf j}_{L_2L_3}^{\nu_1\nu_2\nu_3}(x)+{\sf j}_{L_3L_2}^{3-\nu_{123},\nu_3\nu_2}(x)\Big)\,,\label{sJ}\\
{\sf j}_{L_2L_3}^{\nu_1\nu_2\nu_3}(x) &=\frac{(-1)^{L_3}\sin(\pi\nu_3)}{4\cos(\pi(\nu_{13}-\frac{ L_{23}^-}{2}))}\frac{\Gamma(1-\nu_1)\Gamma(1-\nu_2)\Gamma(\frac{3}{2}-\nu_2)\Gamma(\frac{3}{2}-\nu_3+\frac{L_{23}^+}{2})}{\Gamma(1-L_2-\nu_2)\Gamma(\nu_3-\frac{L_{23}^-}{2})}\nn
	&\quad\times x^{L_2}\HYPwR{\nu_2-\frac{1}{2}}{\nu_1+L_2}{1-\nu_3+\frac{L_{23}^-}{2}}{\frac{3}{2}-\nu_3+\frac{L_{23}^+}{2}}{\frac{3}{2}+L_2}{\frac{5}{2}-\nu_{13}+\frac{L_{23}^-}{2}}{3-\nu_{13}+\frac{L_{23}^+}{2}}{x^2},\label{jfunc}
\end{align}
\end{widetext}
for $x\le 1$, where $L_{23}^\pm\equiv L_2\pm L_3$ and ${}_{p}\widetilde F_q$ is the regularized generalized hypergeometric function. 
For $x>1$, we can use the inversion formula
\begin{equation}
	\sJ_{L_2L_3}^{\nu_1\nu_2\nu_3}(x)=x^{3-2\nu_{123}}\sJ_{L_2L_3}^{\nu_1\nu_3\nu_2}(x^{-1})
\end{equation}
to evaluate the series using the right-hand side. 
From the representation \eqref{Jint2}, it is straightforward to extract the bispectrum multipoles in redshift space~\cite{Scoccimarro:1999ed,Scoccimarro:2015bla, Yankelevich:2018uaz,Agarwal:2020lov,Gualdi:2020ymf,Pardede:2022udo,Amendola:2023awr,DAmico:2022osl,Philcox:2022frc,Ivanov:2023qzb,Chen:2024vuf}. \\[-5pt]

\noindent{\bf Simple Limits.}
It is instructive to check that our formula reproduces the expected results in certain simple limits.

\vskip 5pt
\noindent{\it Monopole $\ell=0$.} In this case, the tripolar spherical harmonic simply becomes
\begin{align}
	S^{LL' 0}(\hat\k_1,\hat\k_2,\hat\n ) =  \frac{2L+1}{4\pi}\delta_{LL'}\cL_{L}(\hat\k_1\cdot\hat\k_2)\,,\label{SLL0}
\end{align}
where $\delta_{LL'}$ is the Kronecker delta. The master integral can then be expressed as
\begin{align}
	&J_{\ell=0}^{\nu_1\nu_2\nu_3}(\k_1,\k_2,\hat\n) \\
	&\quad= k_1^{3-2\nu_{123}}\sum_{L=0}^\infty \frac{2L+1}{4\pi} \sJ_{LL}^{\nu_1\nu_2\nu_3}(x)\cL_L(\hat\k_1\cdot\hat\k_2)\,,\nonumber\label{Jintl0}
\end{align}
which is precisely in the form~\eqref{proj} for a function of the angle $\hat\k_1\cdot\hat\k_2$.

\vskip 5pt
\noindent{\it Power spectrum $\nu_3=0$.} 
Setting $\nu_3=L_3=0$, the phase factor $\sin(\pi\nu_3)$ vanishes for the first term in \eqref{sJ}. 
In addition, one of the parameters of the $_4F_3$ function becomes zero, so that $_4F_3[\cdots]=1$. 
The remaining gamma functions then precisely reduce to the one-loop power spectrum integral $\sI_L$, giving $\sJ_{L0}^{\nu_1\nu_2 0} (x) = 4\pi \sI_{L}^{\nu_1\nu_2}$. 
Combining with~\eqref{SLL0}, we find that  $J_{\ell}^{\nu_1\nu_2 0}(\k_1,\k_2,\hat\n) =I_\ell^{\nu_1\nu_2}(\hat\k_1\cdot\hat\n)$, as expected.\\

\noindent{\bf Convergence.}
Let us comment on the convergence of the expressions. 
There are two types of convergence to consider: $i$) the series expansion of the hypergeometric function in \eqref{jfunc} and $ii$) the multipole sum in \eqref{Jint2}.
Note that the former just depends on the ratio of the two momentum magnitudes, not on the full shape of the triangle. 
The hypergeometric series always converges for $x<1$ and converges faster for smaller $x$.
To optimize the convergence rate, we can choose $x$ to be the ratio of the shortest and longest sides of the triangle, i.e., $x=\frac{\min(k_1,k_2,k_3)}{\max(k_1,k_2,k_3)}$. 
For $x\ll 1$, the series converges rapidly, requiring only a few terms.

The radial integral scales as $L^{\nu_{23}-5/2}x^L$ for large $L$. 
The limiting case is the equilateral configuration $k_1=k_2=k_3$, for which $x=1$. 
In this case, the hypergeometric series in \eqref{jfunc} only converges when $\Re[5-2\nu_{23}]>0$. 
It turns out that this condition is not always met, and even when it is met the convergence can be rather slow for $x\approx 1$. 
As described in~\cite{Simonovic:2017mhp}, this convergence condition can be met by lowering the parameters $\nu_2$ or $\nu_3$ using recursion relations derived from integration-by-parts identities.

The advantage of our approach is that it avoids numerical instability associated with the artificial splitting of tensor structures, each of which can exhibit divergences in specific kinematic configurations~\cite{Philcox:2022frc}. 
Instead, we have organized the tensor structures using spherically symmetric functions~\eqref{TSH}, with coefficients represented by a single scalar integral for arbitrary multipoles, which is free of spurious divergences.
Since our series expansion is intrinsically defined in the squeezed limit, the scalar integrals may converge somewhat slowly in the equilateral limit.
For practical implementation, an optimal strategy may involve a hybrid approach: using rotational symmetry to organize the tensor structures, while employing more optimal representations of the scalar integrals depending on the kinematic region.

\section{Conclusions}

The coming decades promise to be an exciting period for LSS studies, with a wealth of new available data.
However, efficiently computing LSS integrals remains challenging, even at subleading, one-loop order in perturbation theory.
As such, advancing analytic approaches for these integrals is important for more precise and accurate understanding of the LSS.

In this {\it letter}, we have presented a new method for computing LSS loop integrals in redshift space.
Specifically, we derived new analytic expressions for the master integrals in the FFTLog basis at one-loop order.
These integrals are formally equivalent to tensor integrals of massless QFT loop amplitudes in three dimensions, with complex exponents of the propagators. 
By exploiting the simplicity of angular integration on $S^2$, we derived analytic expressions for the master integrals for the one-loop power spectrum and bispectrum in redshift space, valid for all multipole moments. 
In Appendix~\ref{app:deriv}, we further provide results for the one-loop trispectrum in redshift space.  
A full implementation of our results in numerical pipelines, including various bias operators, and the generalization to two-loop order are left to future work.

A central advantage of the FFTLog-based method of~\cite{Simonovic:2017mhp} is the complete factorization of the cosmology dependence from the integrals, allowing them to be computed once and recycled for many cosmologies. 
However, depending on the desired precision and kinematic regions of interest, users may still need to evaluate the master integrals for different choices of $\nu$ parameters in those specific regions.
Therefore, having more efficient representations for evaluation is beneficial, even when the integrals have been precomputed for an earlier analysis.
We hope that our results can help accelerate the analysis of the upcoming LSS data.

While we specifically studied the LSS tensor integrals in the FFTLog basis of scaling universes, the techniques of spherical projection using the Funk-Hecke formula are general.
Most immediately, our approach can be applied to LSS tensor integrals in other bases, such as the basis of massive propagators used in~\cite{Anastasiou:2022udy}.
More generally, similar angular integrals also arise in particle physics, including phase-space integrals for cross sections~\cite{Somogyi:2011ir,Lyubovitskij:2021ges} and helicity projections in conformal field theory~\cite{Isono:2019wex}. 
It would be interesting to explore applications of our approach for computing tensor and angular integrals in these broader contexts.

\vskip 5pt
\begin{acknowledgments}

\noindent{\bf  Acknowledgments.}~We thank Daniel Baumann, Giovanni Cabass, John Joseph Carrasco, Lance Dixon, Daniel Green, Wayne Hu, Donghui Jeong, Austin Joyce, Marko Simonovi\'c, and Mikhail Solon for useful discussions, as well as Matthew Lewandowski and Mikhail Ivanov for comments on the draft.
This work was supported by the Kavli Institute for Cosmological Physics.
\end{acknowledgments}

\bibliographystyle{apsrev4-2}
\bibliography{lssintegral}

\begin{thebibliography}{90}%
\makeatletter
\providecommand \@ifxundefined [1]{%
 \@ifx{#1\undefined}
}%
\providecommand \@ifnum [1]{%
 \ifnum #1\expandafter \@firstoftwo
 \else \expandafter \@secondoftwo
 \fi
}%
\providecommand \@ifx [1]{%
 \ifx #1\expandafter \@firstoftwo
 \else \expandafter \@secondoftwo
 \fi
}%
\providecommand \natexlab [1]{#1}%
\providecommand \enquote  [1]{``#1''}%
\providecommand \bibnamefont  [1]{#1}%
\providecommand \bibfnamefont [1]{#1}%
\providecommand \citenamefont [1]{#1}%
\providecommand \href@noop [0]{\@secondoftwo}%
\providecommand \href [0]{\begingroup \@sanitize@url \@href}%
\providecommand \@href[1]{\@@startlink{#1}\@@href}%
\providecommand \@@href[1]{\endgroup#1\@@endlink}%
\providecommand \@sanitize@url [0]{\catcode `\\12\catcode `\$12\catcode
  `\&12\catcode `\#12\catcode `\^12\catcode `\_12\catcode `\%12\relax}%
\providecommand \@@startlink[1]{}%
\providecommand \@@endlink[0]{}%
\providecommand \url  [0]{\begingroup\@sanitize@url \@url }%
\providecommand \@url [1]{\endgroup\@href {#1}{\urlprefix }}%
\providecommand \urlprefix  [0]{URL }%
\providecommand \Eprint [0]{\href }%
\providecommand \doibase [0]{https://doi.org/}%
\providecommand \selectlanguage [0]{\@gobble}%
\providecommand \bibinfo  [0]{\@secondoftwo}%
\providecommand \bibfield  [0]{\@secondoftwo}%
\providecommand \translation [1]{[#1]}%
\providecommand \BibitemOpen [0]{}%
\providecommand \bibitemStop [0]{}%
\providecommand \bibitemNoStop [0]{.\EOS\space}%
\providecommand \EOS [0]{\spacefactor3000\relax}%
\providecommand \BibitemShut  [1]{\csname bibitem#1\endcsname}%
\let\auto@bib@innerbib\@empty
\bibitem [{\citenamefont {Aghamousa}\ \emph {et~al.}(2016)\citenamefont
  {Aghamousa} \emph {et~al.}}]{DESI:2016fyo}%
  \BibitemOpen
  \bibfield  {author} {\bibinfo {author} {\bibfnamefont {A.}~\bibnamefont
  {Aghamousa}} \emph {et~al.} (\bibinfo {collaboration} {DESI}),\ }\href@noop
  {} {\  (\bibinfo {year} {2016})},\ \Eprint {https://arxiv.org/abs/1611.00036}
  {arXiv:1611.00036 [astro-ph.IM]} \BibitemShut {NoStop}%
\bibitem [{\citenamefont {Laureijs}\ \emph {et~al.}(2011)\citenamefont
  {Laureijs} \emph {et~al.}}]{EUCLID:2011zbd}%
  \BibitemOpen
  \bibfield  {author} {\bibinfo {author} {\bibfnamefont {R.}~\bibnamefont
  {Laureijs}} \emph {et~al.} (\bibinfo {collaboration} {EUCLID}),\ }\href@noop
  {} {\  (\bibinfo {year} {2011})},\ \Eprint {https://arxiv.org/abs/1110.3193}
  {arXiv:1110.3193 [astro-ph.CO]} \BibitemShut {NoStop}%
\bibitem [{\citenamefont {Dor\'e}\ \emph {et~al.}(2014)\citenamefont {Dor\'e}
  \emph {et~al.}}]{SPHEREx:2014bgr}%
  \BibitemOpen
  \bibfield  {author} {\bibinfo {author} {\bibfnamefont {O.}~\bibnamefont
  {Dor\'e}} \emph {et~al.} (\bibinfo {collaboration} {SPHEREx}),\ }\href@noop
  {} {\  (\bibinfo {year} {2014})},\ \Eprint {https://arxiv.org/abs/1412.4872}
  {arXiv:1412.4872 [astro-ph.CO]} \BibitemShut {NoStop}%
\bibitem [{\citenamefont {Ivezi\'c}\ \emph {et~al.}(2019)\citenamefont
  {Ivezi\'c} \emph {et~al.}}]{LSST:2008ijt}%
  \BibitemOpen
  \bibfield  {author} {\bibinfo {author} {\bibfnamefont {Z.}~\bibnamefont
  {Ivezi\'c}} \emph {et~al.} (\bibinfo {collaboration} {LSST}),\ }\href
  {https://doi.org/10.3847/1538-4357/ab042c} {\bibfield  {journal} {\bibinfo
  {journal} {Astrophys. J.}\ }\textbf {\bibinfo {volume} {873}},\ \bibinfo
  {pages} {111} (\bibinfo {year} {2019})},\ \Eprint
  {https://arxiv.org/abs/0805.2366} {arXiv:0805.2366 [astro-ph]} \BibitemShut
  {NoStop}%
\bibitem [{\citenamefont {Akeson}\ \emph {et~al.}(2019)\citenamefont {Akeson}
  \emph {et~al.}}]{Akeson:2019biv}%
  \BibitemOpen
  \bibfield  {author} {\bibinfo {author} {\bibfnamefont {R.}~\bibnamefont
  {Akeson}} \emph {et~al.},\ }\href@noop {} {\  (\bibinfo {year} {2019})},\
  \Eprint {https://arxiv.org/abs/1902.05569} {arXiv:1902.05569 [astro-ph.IM]}
  \BibitemShut {NoStop}%
\bibitem [{\citenamefont {Schlegel}\ \emph {et~al.}(2022)\citenamefont
  {Schlegel} \emph {et~al.}}]{Schlegel:2022vrv}%
  \BibitemOpen
  \bibfield  {author} {\bibinfo {author} {\bibfnamefont {D.~J.}\ \bibnamefont
  {Schlegel}} \emph {et~al.},\ }\href@noop {} {\  (\bibinfo {year} {2022})},\
  \Eprint {https://arxiv.org/abs/2209.04322} {arXiv:2209.04322 [astro-ph.IM]}
  \BibitemShut {NoStop}%
\bibitem [{\citenamefont {Baumann}\ \emph {et~al.}(2012)\citenamefont
  {Baumann}, \citenamefont {Nicolis}, \citenamefont {Senatore},\ and\
  \citenamefont {Zaldarriaga}}]{Baumann:2010tm}%
  \BibitemOpen
  \bibfield  {author} {\bibinfo {author} {\bibfnamefont {D.}~\bibnamefont
  {Baumann}}, \bibinfo {author} {\bibfnamefont {A.}~\bibnamefont {Nicolis}},
  \bibinfo {author} {\bibfnamefont {L.}~\bibnamefont {Senatore}},\ and\
  \bibinfo {author} {\bibfnamefont {M.}~\bibnamefont {Zaldarriaga}},\ }\href
  {https://doi.org/10.1088/1475-7516/2012/07/051} {\bibfield  {journal}
  {\bibinfo  {journal} {JCAP}\ }\textbf {\bibinfo {volume} {07}},\ \bibinfo
  {pages} {051}},\ \Eprint {https://arxiv.org/abs/1004.2488} {arXiv:1004.2488
  [astro-ph.CO]} \BibitemShut {NoStop}%
\bibitem [{\citenamefont {Carrasco}\ \emph {et~al.}(2012)\citenamefont
  {Carrasco}, \citenamefont {Hertzberg},\ and\ \citenamefont
  {Senatore}}]{Carrasco:2012cv}%
  \BibitemOpen
  \bibfield  {author} {\bibinfo {author} {\bibfnamefont {J.~J.~M.}\
  \bibnamefont {Carrasco}}, \bibinfo {author} {\bibfnamefont {M.~P.}\
  \bibnamefont {Hertzberg}},\ and\ \bibinfo {author} {\bibfnamefont
  {L.}~\bibnamefont {Senatore}},\ }\href
  {https://doi.org/10.1007/JHEP09(2012)082} {\bibfield  {journal} {\bibinfo
  {journal} {JHEP}\ }\textbf {\bibinfo {volume} {09}},\ \bibinfo {pages}
  {082}},\ \Eprint {https://arxiv.org/abs/1206.2926} {arXiv:1206.2926
  [astro-ph.CO]} \BibitemShut {NoStop}%
\bibitem [{\citenamefont {Cabass}\ \emph {et~al.}(2023)\citenamefont {Cabass},
  \citenamefont {Ivanov}, \citenamefont {Lewandowski}, \citenamefont
  {Mirbabayi},\ and\ \citenamefont {Simonovi\'c}}]{Cabass:2022avo}%
  \BibitemOpen
  \bibfield  {author} {\bibinfo {author} {\bibfnamefont {G.}~\bibnamefont
  {Cabass}}, \bibinfo {author} {\bibfnamefont {M.~M.}\ \bibnamefont {Ivanov}},
  \bibinfo {author} {\bibfnamefont {M.}~\bibnamefont {Lewandowski}}, \bibinfo
  {author} {\bibfnamefont {M.}~\bibnamefont {Mirbabayi}},\ and\ \bibinfo
  {author} {\bibfnamefont {M.}~\bibnamefont {Simonovi\'c}},\ }\href
  {https://doi.org/10.1016/j.dark.2023.101193} {\bibfield  {journal} {\bibinfo
  {journal} {Phys. Dark Univ.}\ }\textbf {\bibinfo {volume} {40}},\ \bibinfo
  {pages} {101193} (\bibinfo {year} {2023})},\ \Eprint
  {https://arxiv.org/abs/2203.08232} {arXiv:2203.08232 [astro-ph.CO]}
  \BibitemShut {NoStop}%
\bibitem [{\citenamefont {Ivanov}(2023)}]{Ivanov:2022mrd}%
  \BibitemOpen
  \bibfield  {author} {\bibinfo {author} {\bibfnamefont {M.~M.}\ \bibnamefont
  {Ivanov}},\ }\bibinfo {title} {{Effective Field Theory for Large-Scale
  Structure}}\ (\bibinfo {year} {2023})\ \Eprint
  {https://arxiv.org/abs/2212.08488} {arXiv:2212.08488 [astro-ph.CO]}
  \BibitemShut {NoStop}%
\bibitem [{\citenamefont {Carrasco}\ \emph
  {et~al.}(2014{\natexlab{a}})\citenamefont {Carrasco}, \citenamefont
  {Foreman}, \citenamefont {Green},\ and\ \citenamefont
  {Senatore}}]{Carrasco:2013sva}%
  \BibitemOpen
  \bibfield  {author} {\bibinfo {author} {\bibfnamefont {J.~J.~M.}\
  \bibnamefont {Carrasco}}, \bibinfo {author} {\bibfnamefont {S.}~\bibnamefont
  {Foreman}}, \bibinfo {author} {\bibfnamefont {D.}~\bibnamefont {Green}},\
  and\ \bibinfo {author} {\bibfnamefont {L.}~\bibnamefont {Senatore}},\ }\href
  {https://doi.org/10.1088/1475-7516/2014/07/056} {\bibfield  {journal}
  {\bibinfo  {journal} {JCAP}\ }\textbf {\bibinfo {volume} {07}},\ \bibinfo
  {pages} {056}},\ \Eprint {https://arxiv.org/abs/1304.4946} {arXiv:1304.4946
  [astro-ph.CO]} \BibitemShut {NoStop}%
\bibitem [{\citenamefont {Carrasco}\ \emph
  {et~al.}(2014{\natexlab{b}})\citenamefont {Carrasco}, \citenamefont
  {Foreman}, \citenamefont {Green},\ and\ \citenamefont
  {Senatore}}]{Carrasco:2013mua}%
  \BibitemOpen
  \bibfield  {author} {\bibinfo {author} {\bibfnamefont {J.~J.~M.}\
  \bibnamefont {Carrasco}}, \bibinfo {author} {\bibfnamefont {S.}~\bibnamefont
  {Foreman}}, \bibinfo {author} {\bibfnamefont {D.}~\bibnamefont {Green}},\
  and\ \bibinfo {author} {\bibfnamefont {L.}~\bibnamefont {Senatore}},\ }\href
  {https://doi.org/10.1088/1475-7516/2014/07/057} {\bibfield  {journal}
  {\bibinfo  {journal} {JCAP}\ }\textbf {\bibinfo {volume} {07}},\ \bibinfo
  {pages} {057}},\ \Eprint {https://arxiv.org/abs/1310.0464} {arXiv:1310.0464
  [astro-ph.CO]} \BibitemShut {NoStop}%
\bibitem [{\citenamefont {Pajer}\ and\ \citenamefont
  {Zaldarriaga}(2013)}]{Pajer:2013jj}%
  \BibitemOpen
  \bibfield  {author} {\bibinfo {author} {\bibfnamefont {E.}~\bibnamefont
  {Pajer}}\ and\ \bibinfo {author} {\bibfnamefont {M.}~\bibnamefont
  {Zaldarriaga}},\ }\href {https://doi.org/10.1088/1475-7516/2013/08/037}
  {\bibfield  {journal} {\bibinfo  {journal} {JCAP}\ }\textbf {\bibinfo
  {volume} {08}},\ \bibinfo {pages} {037}},\ \Eprint
  {https://arxiv.org/abs/1301.7182} {arXiv:1301.7182 [astro-ph.CO]}
  \BibitemShut {NoStop}%
\bibitem [{\citenamefont {Senatore}\ and\ \citenamefont
  {Zaldarriaga}(2015)}]{Senatore:2014via}%
  \BibitemOpen
  \bibfield  {author} {\bibinfo {author} {\bibfnamefont {L.}~\bibnamefont
  {Senatore}}\ and\ \bibinfo {author} {\bibfnamefont {M.}~\bibnamefont
  {Zaldarriaga}},\ }\href {https://doi.org/10.1088/1475-7516/2015/02/013}
  {\bibfield  {journal} {\bibinfo  {journal} {JCAP}\ }\textbf {\bibinfo
  {volume} {02}},\ \bibinfo {pages} {013}},\ \Eprint
  {https://arxiv.org/abs/1404.5954} {arXiv:1404.5954 [astro-ph.CO]}
  \BibitemShut {NoStop}%
\bibitem [{\citenamefont {Baldauf}\ \emph
  {et~al.}(2015{\natexlab{a}})\citenamefont {Baldauf}, \citenamefont
  {Mercolli},\ and\ \citenamefont {Zaldarriaga}}]{Baldauf:2015aha}%
  \BibitemOpen
  \bibfield  {author} {\bibinfo {author} {\bibfnamefont {T.}~\bibnamefont
  {Baldauf}}, \bibinfo {author} {\bibfnamefont {L.}~\bibnamefont {Mercolli}},\
  and\ \bibinfo {author} {\bibfnamefont {M.}~\bibnamefont {Zaldarriaga}},\
  }\href {https://doi.org/10.1103/PhysRevD.92.123007} {\bibfield  {journal}
  {\bibinfo  {journal} {Phys. Rev. D}\ }\textbf {\bibinfo {volume} {92}},\
  \bibinfo {pages} {123007} (\bibinfo {year} {2015}{\natexlab{a}})},\ \Eprint
  {https://arxiv.org/abs/1507.02256} {arXiv:1507.02256 [astro-ph.CO]}
  \BibitemShut {NoStop}%
\bibitem [{\citenamefont {Konstandin}\ \emph {et~al.}(2019)\citenamefont
  {Konstandin}, \citenamefont {Porto},\ and\ \citenamefont
  {Rubira}}]{Konstandin:2019bay}%
  \BibitemOpen
  \bibfield  {author} {\bibinfo {author} {\bibfnamefont {T.}~\bibnamefont
  {Konstandin}}, \bibinfo {author} {\bibfnamefont {R.~A.}\ \bibnamefont
  {Porto}},\ and\ \bibinfo {author} {\bibfnamefont {H.}~\bibnamefont
  {Rubira}},\ }\href {https://doi.org/10.1088/1475-7516/2019/11/027} {\bibfield
   {journal} {\bibinfo  {journal} {JCAP}\ }\textbf {\bibinfo {volume} {11}},\
  \bibinfo {pages} {027}},\ \Eprint {https://arxiv.org/abs/1906.00997}
  {arXiv:1906.00997 [astro-ph.CO]} \BibitemShut {NoStop}%
\bibitem [{\citenamefont {Angulo}\ \emph {et~al.}(2015)\citenamefont {Angulo},
  \citenamefont {Foreman}, \citenamefont {Schmittfull},\ and\ \citenamefont
  {Senatore}}]{Angulo:2014tfa}%
  \BibitemOpen
  \bibfield  {author} {\bibinfo {author} {\bibfnamefont {R.~E.}\ \bibnamefont
  {Angulo}}, \bibinfo {author} {\bibfnamefont {S.}~\bibnamefont {Foreman}},
  \bibinfo {author} {\bibfnamefont {M.}~\bibnamefont {Schmittfull}},\ and\
  \bibinfo {author} {\bibfnamefont {L.}~\bibnamefont {Senatore}},\ }\href
  {https://doi.org/10.1088/1475-7516/2015/10/039} {\bibfield  {journal}
  {\bibinfo  {journal} {JCAP}\ }\textbf {\bibinfo {volume} {10}},\ \bibinfo
  {pages} {039}},\ \Eprint {https://arxiv.org/abs/1406.4143} {arXiv:1406.4143
  [astro-ph.CO]} \BibitemShut {NoStop}%
\bibitem [{\citenamefont {Baldauf}\ \emph
  {et~al.}(2015{\natexlab{b}})\citenamefont {Baldauf}, \citenamefont
  {Mercolli}, \citenamefont {Mirbabayi},\ and\ \citenamefont
  {Pajer}}]{Baldauf:2014qfa}%
  \BibitemOpen
  \bibfield  {author} {\bibinfo {author} {\bibfnamefont {T.}~\bibnamefont
  {Baldauf}}, \bibinfo {author} {\bibfnamefont {L.}~\bibnamefont {Mercolli}},
  \bibinfo {author} {\bibfnamefont {M.}~\bibnamefont {Mirbabayi}},\ and\
  \bibinfo {author} {\bibfnamefont {E.}~\bibnamefont {Pajer}},\ }\href
  {https://doi.org/10.1088/1475-7516/2015/05/007} {\bibfield  {journal}
  {\bibinfo  {journal} {JCAP}\ }\textbf {\bibinfo {volume} {05}},\ \bibinfo
  {pages} {007}},\ \Eprint {https://arxiv.org/abs/1406.4135} {arXiv:1406.4135
  [astro-ph.CO]} \BibitemShut {NoStop}%
\bibitem [{\citenamefont {Eggemeier}\ \emph {et~al.}(2019)\citenamefont
  {Eggemeier}, \citenamefont {Scoccimarro},\ and\ \citenamefont
  {Smith}}]{Eggemeier:2018qae}%
  \BibitemOpen
  \bibfield  {author} {\bibinfo {author} {\bibfnamefont {A.}~\bibnamefont
  {Eggemeier}}, \bibinfo {author} {\bibfnamefont {R.}~\bibnamefont
  {Scoccimarro}},\ and\ \bibinfo {author} {\bibfnamefont {R.~E.}\ \bibnamefont
  {Smith}},\ }\href {https://doi.org/10.1103/PhysRevD.99.123514} {\bibfield
  {journal} {\bibinfo  {journal} {Phys. Rev. D}\ }\textbf {\bibinfo {volume}
  {99}},\ \bibinfo {pages} {123514} (\bibinfo {year} {2019})},\ \Eprint
  {https://arxiv.org/abs/1812.03208} {arXiv:1812.03208 [astro-ph.CO]}
  \BibitemShut {NoStop}%
\bibitem [{\citenamefont {D'Amico}\ \emph {et~al.}(2022)\citenamefont
  {D'Amico}, \citenamefont {Donath}, \citenamefont {Lewandowski}, \citenamefont
  {Senatore},\ and\ \citenamefont {Zhang}}]{DAmico:2022ukl}%
  \BibitemOpen
  \bibfield  {author} {\bibinfo {author} {\bibfnamefont {G.}~\bibnamefont
  {D'Amico}}, \bibinfo {author} {\bibfnamefont {Y.}~\bibnamefont {Donath}},
  \bibinfo {author} {\bibfnamefont {M.}~\bibnamefont {Lewandowski}}, \bibinfo
  {author} {\bibfnamefont {L.}~\bibnamefont {Senatore}},\ and\ \bibinfo
  {author} {\bibfnamefont {P.}~\bibnamefont {Zhang}},\ }\href@noop {} {\
  (\bibinfo {year} {2022})},\ \Eprint {https://arxiv.org/abs/2211.17130}
  {arXiv:2211.17130 [astro-ph.CO]} \BibitemShut {NoStop}%
\bibitem [{\citenamefont {Bertolini}\ \emph
  {et~al.}(2016{\natexlab{a}})\citenamefont {Bertolini}, \citenamefont
  {Schutz}, \citenamefont {Solon}, \citenamefont {Walsh},\ and\ \citenamefont
  {Zurek}}]{Bertolini:2015fya}%
  \BibitemOpen
  \bibfield  {author} {\bibinfo {author} {\bibfnamefont {D.}~\bibnamefont
  {Bertolini}}, \bibinfo {author} {\bibfnamefont {K.}~\bibnamefont {Schutz}},
  \bibinfo {author} {\bibfnamefont {M.~P.}\ \bibnamefont {Solon}}, \bibinfo
  {author} {\bibfnamefont {J.~R.}\ \bibnamefont {Walsh}},\ and\ \bibinfo
  {author} {\bibfnamefont {K.~M.}\ \bibnamefont {Zurek}},\ }\href
  {https://doi.org/10.1103/PhysRevD.93.123505} {\bibfield  {journal} {\bibinfo
  {journal} {Phys. Rev. D}\ }\textbf {\bibinfo {volume} {93}},\ \bibinfo
  {pages} {123505} (\bibinfo {year} {2016}{\natexlab{a}})},\ \Eprint
  {https://arxiv.org/abs/1512.07630} {arXiv:1512.07630 [astro-ph.CO]}
  \BibitemShut {NoStop}%
\bibitem [{\citenamefont {Bertolini}\ \emph
  {et~al.}(2016{\natexlab{b}})\citenamefont {Bertolini}, \citenamefont
  {Schutz}, \citenamefont {Solon},\ and\ \citenamefont
  {Zurek}}]{Bertolini:2016bmt}%
  \BibitemOpen
  \bibfield  {author} {\bibinfo {author} {\bibfnamefont {D.}~\bibnamefont
  {Bertolini}}, \bibinfo {author} {\bibfnamefont {K.}~\bibnamefont {Schutz}},
  \bibinfo {author} {\bibfnamefont {M.~P.}\ \bibnamefont {Solon}},\ and\
  \bibinfo {author} {\bibfnamefont {K.~M.}\ \bibnamefont {Zurek}},\ }\href
  {https://doi.org/10.1088/1475-7516/2016/06/052} {\bibfield  {journal}
  {\bibinfo  {journal} {JCAP}\ }\textbf {\bibinfo {volume} {06}},\ \bibinfo
  {pages} {052}},\ \Eprint {https://arxiv.org/abs/1604.01770} {arXiv:1604.01770
  [astro-ph.CO]} \BibitemShut {NoStop}%
\bibitem [{\citenamefont {Steele}\ and\ \citenamefont
  {Baldauf}(2021)}]{Steele:2021lnz}%
  \BibitemOpen
  \bibfield  {author} {\bibinfo {author} {\bibfnamefont {T.}~\bibnamefont
  {Steele}}\ and\ \bibinfo {author} {\bibfnamefont {T.}~\bibnamefont
  {Baldauf}},\ }\href {https://doi.org/10.1103/PhysRevD.103.103518} {\bibfield
  {journal} {\bibinfo  {journal} {Phys. Rev. D}\ }\textbf {\bibinfo {volume}
  {103}},\ \bibinfo {pages} {103518} (\bibinfo {year} {2021})},\ \Eprint
  {https://arxiv.org/abs/2101.10289} {arXiv:2101.10289 [astro-ph.CO]}
  \BibitemShut {NoStop}%
\bibitem [{\citenamefont {Simonovi\'c}\ \emph {et~al.}(2018)\citenamefont
  {Simonovi\'c}, \citenamefont {Baldauf}, \citenamefont {Zaldarriaga},
  \citenamefont {Carrasco},\ and\ \citenamefont
  {Kollmeier}}]{Simonovic:2017mhp}%
  \BibitemOpen
  \bibfield  {author} {\bibinfo {author} {\bibfnamefont {M.}~\bibnamefont
  {Simonovi\'c}}, \bibinfo {author} {\bibfnamefont {T.}~\bibnamefont
  {Baldauf}}, \bibinfo {author} {\bibfnamefont {M.}~\bibnamefont
  {Zaldarriaga}}, \bibinfo {author} {\bibfnamefont {J.~J.}\ \bibnamefont
  {Carrasco}},\ and\ \bibinfo {author} {\bibfnamefont {J.~A.}\ \bibnamefont
  {Kollmeier}},\ }\href {https://doi.org/10.1088/1475-7516/2018/04/030}
  {\bibfield  {journal} {\bibinfo  {journal} {JCAP}\ }\textbf {\bibinfo
  {volume} {04}},\ \bibinfo {pages} {030}},\ \Eprint
  {https://arxiv.org/abs/1708.08130} {arXiv:1708.08130 [astro-ph.CO]}
  \BibitemShut {NoStop}%
\bibitem [{\citenamefont {Anastasiou}\ \emph {et~al.}(2024)\citenamefont
  {Anastasiou}, \citenamefont {Bragan\c{c}a}, \citenamefont {Senatore},\ and\
  \citenamefont {Zheng}}]{Anastasiou:2022udy}%
  \BibitemOpen
  \bibfield  {author} {\bibinfo {author} {\bibfnamefont {C.}~\bibnamefont
  {Anastasiou}}, \bibinfo {author} {\bibfnamefont {D.~P.~L.}\ \bibnamefont
  {Bragan\c{c}a}}, \bibinfo {author} {\bibfnamefont {L.}~\bibnamefont
  {Senatore}},\ and\ \bibinfo {author} {\bibfnamefont {H.}~\bibnamefont
  {Zheng}},\ }\href {https://doi.org/10.1007/JHEP01(2024)002} {\bibfield
  {journal} {\bibinfo  {journal} {JHEP}\ }\textbf {\bibinfo {volume} {01}},\
  \bibinfo {pages} {002}},\ \Eprint {https://arxiv.org/abs/2212.07421}
  {arXiv:2212.07421 [astro-ph.CO]} \BibitemShut {NoStop}%
\bibitem [{\citenamefont {Schmittfull}\ \emph {et~al.}(2016)\citenamefont
  {Schmittfull}, \citenamefont {Vlah},\ and\ \citenamefont
  {McDonald}}]{Schmittfull:2016jsw}%
  \BibitemOpen
  \bibfield  {author} {\bibinfo {author} {\bibfnamefont {M.}~\bibnamefont
  {Schmittfull}}, \bibinfo {author} {\bibfnamefont {Z.}~\bibnamefont {Vlah}},\
  and\ \bibinfo {author} {\bibfnamefont {P.}~\bibnamefont {McDonald}},\ }\href
  {https://doi.org/10.1103/PhysRevD.93.103528} {\bibfield  {journal} {\bibinfo
  {journal} {Phys. Rev. D}\ }\textbf {\bibinfo {volume} {93}},\ \bibinfo
  {pages} {103528} (\bibinfo {year} {2016})},\ \Eprint
  {https://arxiv.org/abs/1603.04405} {arXiv:1603.04405 [astro-ph.CO]}
  \BibitemShut {NoStop}%
\bibitem [{\citenamefont {McEwen}\ \emph {et~al.}(2016)\citenamefont {McEwen},
  \citenamefont {Fang}, \citenamefont {Hirata},\ and\ \citenamefont
  {Blazek}}]{McEwen:2016fjn}%
  \BibitemOpen
  \bibfield  {author} {\bibinfo {author} {\bibfnamefont {J.~E.}\ \bibnamefont
  {McEwen}}, \bibinfo {author} {\bibfnamefont {X.}~\bibnamefont {Fang}},
  \bibinfo {author} {\bibfnamefont {C.~M.}\ \bibnamefont {Hirata}},\ and\
  \bibinfo {author} {\bibfnamefont {J.~A.}\ \bibnamefont {Blazek}},\ }\href
  {https://doi.org/10.1088/1475-7516/2016/09/015} {\bibfield  {journal}
  {\bibinfo  {journal} {JCAP}\ }\textbf {\bibinfo {volume} {09}},\ \bibinfo
  {pages} {015}},\ \Eprint {https://arxiv.org/abs/1603.04826} {arXiv:1603.04826
  [astro-ph.CO]} \BibitemShut {NoStop}%
\bibitem [{\citenamefont {Schmittfull}\ and\ \citenamefont
  {Vlah}(2016)}]{Schmittfull:2016yqx}%
  \BibitemOpen
  \bibfield  {author} {\bibinfo {author} {\bibfnamefont {M.}~\bibnamefont
  {Schmittfull}}\ and\ \bibinfo {author} {\bibfnamefont {Z.}~\bibnamefont
  {Vlah}},\ }\href {https://doi.org/10.1103/PhysRevD.94.103530} {\bibfield
  {journal} {\bibinfo  {journal} {Phys. Rev. D}\ }\textbf {\bibinfo {volume}
  {94}},\ \bibinfo {pages} {103530} (\bibinfo {year} {2016})},\ \Eprint
  {https://arxiv.org/abs/1609.00349} {arXiv:1609.00349 [astro-ph.CO]}
  \BibitemShut {NoStop}%
\bibitem [{\citenamefont {Fang}\ \emph {et~al.}(2017)\citenamefont {Fang},
  \citenamefont {Blazek}, \citenamefont {McEwen},\ and\ \citenamefont
  {Hirata}}]{Fang:2016wcf}%
  \BibitemOpen
  \bibfield  {author} {\bibinfo {author} {\bibfnamefont {X.}~\bibnamefont
  {Fang}}, \bibinfo {author} {\bibfnamefont {J.~A.}\ \bibnamefont {Blazek}},
  \bibinfo {author} {\bibfnamefont {J.~E.}\ \bibnamefont {McEwen}},\ and\
  \bibinfo {author} {\bibfnamefont {C.~M.}\ \bibnamefont {Hirata}},\ }\href
  {https://doi.org/10.1088/1475-7516/2017/02/030} {\bibfield  {journal}
  {\bibinfo  {journal} {JCAP}\ }\textbf {\bibinfo {volume} {02}},\ \bibinfo
  {pages} {030}},\ \Eprint {https://arxiv.org/abs/1609.05978} {arXiv:1609.05978
  [astro-ph.CO]} \BibitemShut {NoStop}%
\bibitem [{\citenamefont {Assassi}\ \emph {et~al.}(2017)\citenamefont
  {Assassi}, \citenamefont {Simonovi\'c},\ and\ \citenamefont
  {Zaldarriaga}}]{Assassi:2017lea}%
  \BibitemOpen
  \bibfield  {author} {\bibinfo {author} {\bibfnamefont {V.}~\bibnamefont
  {Assassi}}, \bibinfo {author} {\bibfnamefont {M.}~\bibnamefont
  {Simonovi\'c}},\ and\ \bibinfo {author} {\bibfnamefont {M.}~\bibnamefont
  {Zaldarriaga}},\ }\href {https://doi.org/10.1088/1475-7516/2017/11/054}
  {\bibfield  {journal} {\bibinfo  {journal} {JCAP}\ }\textbf {\bibinfo
  {volume} {11}},\ \bibinfo {pages} {054}},\ \Eprint
  {https://arxiv.org/abs/1705.05022} {arXiv:1705.05022 [astro-ph.CO]}
  \BibitemShut {NoStop}%
\bibitem [{\citenamefont {Grasshorn~Gebhardt}\ and\ \citenamefont
  {Jeong}(2018)}]{GrasshornGebhardt:2017tbv}%
  \BibitemOpen
  \bibfield  {author} {\bibinfo {author} {\bibfnamefont {H.~S.}\ \bibnamefont
  {Grasshorn~Gebhardt}}\ and\ \bibinfo {author} {\bibfnamefont
  {D.}~\bibnamefont {Jeong}},\ }\href
  {https://doi.org/10.1103/PhysRevD.97.023504} {\bibfield  {journal} {\bibinfo
  {journal} {Phys. Rev. D}\ }\textbf {\bibinfo {volume} {97}},\ \bibinfo
  {pages} {023504} (\bibinfo {year} {2018})},\ \Eprint
  {https://arxiv.org/abs/1709.02401} {arXiv:1709.02401 [astro-ph.CO]}
  \BibitemShut {NoStop}%
\bibitem [{\citenamefont {Sch\"oneberg}\ \emph {et~al.}(2018)\citenamefont
  {Sch\"oneberg}, \citenamefont {Simonovi\'c}, \citenamefont {Lesgourgues},\
  and\ \citenamefont {Zaldarriaga}}]{Schoneberg:2018fis}%
  \BibitemOpen
  \bibfield  {author} {\bibinfo {author} {\bibfnamefont {N.}~\bibnamefont
  {Sch\"oneberg}}, \bibinfo {author} {\bibfnamefont {M.}~\bibnamefont
  {Simonovi\'c}}, \bibinfo {author} {\bibfnamefont {J.}~\bibnamefont
  {Lesgourgues}},\ and\ \bibinfo {author} {\bibfnamefont {M.}~\bibnamefont
  {Zaldarriaga}},\ }\href {https://doi.org/10.1088/1475-7516/2018/10/047}
  {\bibfield  {journal} {\bibinfo  {journal} {JCAP}\ }\textbf {\bibinfo
  {volume} {10}},\ \bibinfo {pages} {047}},\ \Eprint
  {https://arxiv.org/abs/1807.09540} {arXiv:1807.09540 [astro-ph.CO]}
  \BibitemShut {NoStop}%
\bibitem [{\citenamefont {Lee}\ and\ \citenamefont
  {Dvorkin}(2020)}]{Lee:2020ebj}%
  \BibitemOpen
  \bibfield  {author} {\bibinfo {author} {\bibfnamefont {H.}~\bibnamefont
  {Lee}}\ and\ \bibinfo {author} {\bibfnamefont {C.}~\bibnamefont {Dvorkin}},\
  }\href {https://doi.org/10.1088/1475-7516/2020/05/044} {\bibfield  {journal}
  {\bibinfo  {journal} {JCAP}\ }\textbf {\bibinfo {volume} {05}},\ \bibinfo
  {pages} {044}},\ \Eprint {https://arxiv.org/abs/2001.00584} {arXiv:2001.00584
  [astro-ph.CO]} \BibitemShut {NoStop}%
\bibitem [{\citenamefont {Chen}\ \emph
  {et~al.}(2021{\natexlab{a}})\citenamefont {Chen}, \citenamefont {Lee},\ and\
  \citenamefont {Dvorkin}}]{Chen:2021vba}%
  \BibitemOpen
  \bibfield  {author} {\bibinfo {author} {\bibfnamefont {S.-F.}\ \bibnamefont
  {Chen}}, \bibinfo {author} {\bibfnamefont {H.}~\bibnamefont {Lee}},\ and\
  \bibinfo {author} {\bibfnamefont {C.}~\bibnamefont {Dvorkin}},\ }\href
  {https://doi.org/10.1088/1475-7516/2021/05/030} {\bibfield  {journal}
  {\bibinfo  {journal} {JCAP}\ }\textbf {\bibinfo {volume} {05}},\ \bibinfo
  {pages} {030}},\ \Eprint {https://arxiv.org/abs/2103.01229} {arXiv:2103.01229
  [astro-ph.CO]} \BibitemShut {NoStop}%
\bibitem [{\citenamefont {Ivanov}\ \emph {et~al.}(2020)\citenamefont {Ivanov},
  \citenamefont {Simonovi\'c},\ and\ \citenamefont
  {Zaldarriaga}}]{Ivanov:2019pdj}%
  \BibitemOpen
  \bibfield  {author} {\bibinfo {author} {\bibfnamefont {M.~M.}\ \bibnamefont
  {Ivanov}}, \bibinfo {author} {\bibfnamefont {M.}~\bibnamefont
  {Simonovi\'c}},\ and\ \bibinfo {author} {\bibfnamefont {M.}~\bibnamefont
  {Zaldarriaga}},\ }\href {https://doi.org/10.1088/1475-7516/2020/05/042}
  {\bibfield  {journal} {\bibinfo  {journal} {JCAP}\ }\textbf {\bibinfo
  {volume} {05}},\ \bibinfo {pages} {042}},\ \Eprint
  {https://arxiv.org/abs/1909.05277} {arXiv:1909.05277 [astro-ph.CO]}
  \BibitemShut {NoStop}%
\bibitem [{\citenamefont {Colas}\ \emph {et~al.}(2020)\citenamefont {Colas},
  \citenamefont {D'amico}, \citenamefont {Senatore}, \citenamefont {Zhang},\
  and\ \citenamefont {Beutler}}]{Colas:2019ret}%
  \BibitemOpen
  \bibfield  {author} {\bibinfo {author} {\bibfnamefont {T.}~\bibnamefont
  {Colas}}, \bibinfo {author} {\bibfnamefont {G.}~\bibnamefont {D'amico}},
  \bibinfo {author} {\bibfnamefont {L.}~\bibnamefont {Senatore}}, \bibinfo
  {author} {\bibfnamefont {P.}~\bibnamefont {Zhang}},\ and\ \bibinfo {author}
  {\bibfnamefont {F.}~\bibnamefont {Beutler}},\ }\href
  {https://doi.org/10.1088/1475-7516/2020/06/001} {\bibfield  {journal}
  {\bibinfo  {journal} {JCAP}\ }\textbf {\bibinfo {volume} {06}},\ \bibinfo
  {pages} {001}},\ \Eprint {https://arxiv.org/abs/1909.07951} {arXiv:1909.07951
  [astro-ph.CO]} \BibitemShut {NoStop}%
\bibitem [{\citenamefont {Chen}\ \emph
  {et~al.}(2021{\natexlab{b}})\citenamefont {Chen}, \citenamefont {Vlah},\ and\
  \citenamefont {White}}]{Chen:2021rnb}%
  \BibitemOpen
  \bibfield  {author} {\bibinfo {author} {\bibfnamefont {S.-F.}\ \bibnamefont
  {Chen}}, \bibinfo {author} {\bibfnamefont {Z.}~\bibnamefont {Vlah}},\ and\
  \bibinfo {author} {\bibfnamefont {M.}~\bibnamefont {White}},\ }\href
  {https://doi.org/10.1088/1475-7516/2021/05/053} {\bibfield  {journal}
  {\bibinfo  {journal} {JCAP}\ }\textbf {\bibinfo {volume} {05}},\ \bibinfo
  {pages} {053}},\ \Eprint {https://arxiv.org/abs/2103.13498} {arXiv:2103.13498
  [astro-ph.CO]} \BibitemShut {NoStop}%
\bibitem [{\citenamefont {Qin}\ \emph {et~al.}(2022)\citenamefont {Qin},
  \citenamefont {Schutz}, \citenamefont {Smith}, \citenamefont {Garaldi},
  \citenamefont {Kannan}, \citenamefont {Slatyer},\ and\ \citenamefont
  {Vogelsberger}}]{Qin:2022xho}%
  \BibitemOpen
  \bibfield  {author} {\bibinfo {author} {\bibfnamefont {W.}~\bibnamefont
  {Qin}}, \bibinfo {author} {\bibfnamefont {K.}~\bibnamefont {Schutz}},
  \bibinfo {author} {\bibfnamefont {A.}~\bibnamefont {Smith}}, \bibinfo
  {author} {\bibfnamefont {E.}~\bibnamefont {Garaldi}}, \bibinfo {author}
  {\bibfnamefont {R.}~\bibnamefont {Kannan}}, \bibinfo {author} {\bibfnamefont
  {T.~R.}\ \bibnamefont {Slatyer}},\ and\ \bibinfo {author} {\bibfnamefont
  {M.}~\bibnamefont {Vogelsberger}},\ }\href
  {https://doi.org/10.1103/PhysRevD.106.123506} {\bibfield  {journal} {\bibinfo
   {journal} {Phys. Rev. D}\ }\textbf {\bibinfo {volume} {106}},\ \bibinfo
  {pages} {123506} (\bibinfo {year} {2022})},\ \Eprint
  {https://arxiv.org/abs/2205.06270} {arXiv:2205.06270 [astro-ph.CO]}
  \BibitemShut {NoStop}%
\bibitem [{\citenamefont {Ivanov}(2024)}]{Ivanov:2023yla}%
  \BibitemOpen
  \bibfield  {author} {\bibinfo {author} {\bibfnamefont {M.~M.}\ \bibnamefont
  {Ivanov}},\ }\href {https://doi.org/10.1103/PhysRevD.109.023507} {\bibfield
  {journal} {\bibinfo  {journal} {Phys. Rev. D}\ }\textbf {\bibinfo {volume}
  {109}},\ \bibinfo {pages} {023507} (\bibinfo {year} {2024})},\ \Eprint
  {https://arxiv.org/abs/2309.10133} {arXiv:2309.10133 [astro-ph.CO]}
  \BibitemShut {NoStop}%
\bibitem [{\citenamefont {Henn}\ and\ \citenamefont
  {Plefka}(2014)}]{Henn:2014yza}%
  \BibitemOpen
  \bibfield  {author} {\bibinfo {author} {\bibfnamefont {J.~M.}\ \bibnamefont
  {Henn}}\ and\ \bibinfo {author} {\bibfnamefont {J.~C.}\ \bibnamefont
  {Plefka}},\ }\href {https://doi.org/10.1007/978-3-642-54022-6} {\emph
  {\bibinfo {title} {{Scattering Amplitudes in Gauge Theories}}}},\ Vol.\
  \bibinfo {volume} {883}\ (\bibinfo  {publisher} {Springer},\ \bibinfo
  {address} {Berlin},\ \bibinfo {year} {2014})\BibitemShut {NoStop}%
\bibitem [{\citenamefont {Weinzierl}(2022)}]{Weinzierl:2022eaz}%
  \BibitemOpen
  \bibfield  {author} {\bibinfo {author} {\bibfnamefont {S.}~\bibnamefont
  {Weinzierl}},\ }\href {https://doi.org/10.1007/978-3-030-99558-4} {\emph
  {\bibinfo {title} {{Feynman Integrals. A Comprehensive Treatment for Students
  and Researchers}}}},\ UNITEXT for Physics\ (\bibinfo  {publisher}
  {Springer},\ \bibinfo {year} {2022})\ \Eprint
  {https://arxiv.org/abs/2201.03593} {arXiv:2201.03593 [hep-th]} \BibitemShut
  {NoStop}%
\bibitem [{\citenamefont {Passarino}\ and\ \citenamefont
  {Veltman}(1979)}]{Passarino:1978jh}%
  \BibitemOpen
  \bibfield  {author} {\bibinfo {author} {\bibfnamefont {G.}~\bibnamefont
  {Passarino}}\ and\ \bibinfo {author} {\bibfnamefont {M.~J.~G.}\ \bibnamefont
  {Veltman}},\ }\href {https://doi.org/10.1016/0550-3213(79)90234-7} {\bibfield
   {journal} {\bibinfo  {journal} {Nucl. Phys. B}\ }\textbf {\bibinfo {volume}
  {160}},\ \bibinfo {pages} {151} (\bibinfo {year} {1979})}\BibitemShut
  {NoStop}%
\bibitem [{\citenamefont {Chudaykin}\ \emph {et~al.}(2020)\citenamefont
  {Chudaykin}, \citenamefont {Ivanov}, \citenamefont {Philcox},\ and\
  \citenamefont {Simonovi\'c}}]{Chudaykin:2020aoj}%
  \BibitemOpen
  \bibfield  {author} {\bibinfo {author} {\bibfnamefont {A.}~\bibnamefont
  {Chudaykin}}, \bibinfo {author} {\bibfnamefont {M.~M.}\ \bibnamefont
  {Ivanov}}, \bibinfo {author} {\bibfnamefont {O.~H.~E.}\ \bibnamefont
  {Philcox}},\ and\ \bibinfo {author} {\bibfnamefont {M.}~\bibnamefont
  {Simonovi\'c}},\ }\href {https://doi.org/10.1103/PhysRevD.102.063533}
  {\bibfield  {journal} {\bibinfo  {journal} {Phys. Rev. D}\ }\textbf {\bibinfo
  {volume} {102}},\ \bibinfo {pages} {063533} (\bibinfo {year} {2020})},\
  \Eprint {https://arxiv.org/abs/2004.10607} {arXiv:2004.10607 [astro-ph.CO]}
  \BibitemShut {NoStop}%
\bibitem [{\citenamefont {Philcox}\ \emph {et~al.}(2022)\citenamefont
  {Philcox}, \citenamefont {Ivanov}, \citenamefont {Cabass}, \citenamefont
  {Simonovi\'c}, \citenamefont {Zaldarriaga},\ and\ \citenamefont
  {Nishimichi}}]{Philcox:2022frc}%
  \BibitemOpen
  \bibfield  {author} {\bibinfo {author} {\bibfnamefont {O.~H.~E.}\
  \bibnamefont {Philcox}}, \bibinfo {author} {\bibfnamefont {M.~M.}\
  \bibnamefont {Ivanov}}, \bibinfo {author} {\bibfnamefont {G.}~\bibnamefont
  {Cabass}}, \bibinfo {author} {\bibfnamefont {M.}~\bibnamefont {Simonovi\'c}},
  \bibinfo {author} {\bibfnamefont {M.}~\bibnamefont {Zaldarriaga}},\ and\
  \bibinfo {author} {\bibfnamefont {T.}~\bibnamefont {Nishimichi}},\ }\href
  {https://doi.org/10.1103/PhysRevD.106.043530} {\bibfield  {journal} {\bibinfo
   {journal} {Phys. Rev. D}\ }\textbf {\bibinfo {volume} {106}},\ \bibinfo
  {pages} {043530} (\bibinfo {year} {2022})},\ \Eprint
  {https://arxiv.org/abs/2206.02800} {arXiv:2206.02800 [astro-ph.CO]}
  \BibitemShut {NoStop}%
\bibitem [{\citenamefont {Davydychev}(1991)}]{Davydychev:1991va}%
  \BibitemOpen
  \bibfield  {author} {\bibinfo {author} {\bibfnamefont {A.~I.}\ \bibnamefont
  {Davydychev}},\ }\href {https://doi.org/10.1016/0370-2693(91)91715-8}
  {\bibfield  {journal} {\bibinfo  {journal} {Phys. Lett. B}\ }\textbf
  {\bibinfo {volume} {263}},\ \bibinfo {pages} {107} (\bibinfo {year}
  {1991})}\BibitemShut {NoStop}%
\bibitem [{\citenamefont {Tarasov}(1996)}]{Tarasov:1996br}%
  \BibitemOpen
  \bibfield  {author} {\bibinfo {author} {\bibfnamefont {O.~V.}\ \bibnamefont
  {Tarasov}},\ }\href {https://doi.org/10.1103/PhysRevD.54.6479} {\bibfield
  {journal} {\bibinfo  {journal} {Phys. Rev. D}\ }\textbf {\bibinfo {volume}
  {54}},\ \bibinfo {pages} {6479} (\bibinfo {year} {1996})},\ \Eprint
  {https://arxiv.org/abs/hep-th/9606018} {arXiv:hep-th/9606018} \BibitemShut
  {NoStop}%
\bibitem [{\citenamefont {Bern}\ \emph {et~al.}(1994)\citenamefont {Bern},
  \citenamefont {Dixon},\ and\ \citenamefont {Kosower}}]{Bern:1993kr}%
  \BibitemOpen
  \bibfield  {author} {\bibinfo {author} {\bibfnamefont {Z.}~\bibnamefont
  {Bern}}, \bibinfo {author} {\bibfnamefont {L.~J.}\ \bibnamefont {Dixon}},\
  and\ \bibinfo {author} {\bibfnamefont {D.~A.}\ \bibnamefont {Kosower}},\
  }\href {https://doi.org/10.1016/0550-3213(94)90398-0} {\bibfield  {journal}
  {\bibinfo  {journal} {Nucl. Phys. B}\ }\textbf {\bibinfo {volume} {412}},\
  \bibinfo {pages} {751} (\bibinfo {year} {1994})},\ \Eprint
  {https://arxiv.org/abs/hep-ph/9306240} {arXiv:hep-ph/9306240} \BibitemShut
  {NoStop}%
\bibitem [{\citenamefont {Binoth}\ \emph {et~al.}(2000)\citenamefont {Binoth},
  \citenamefont {Guillet},\ and\ \citenamefont {Heinrich}}]{Binoth:1999sp}%
  \BibitemOpen
  \bibfield  {author} {\bibinfo {author} {\bibfnamefont {T.}~\bibnamefont
  {Binoth}}, \bibinfo {author} {\bibfnamefont {J.~P.}\ \bibnamefont
  {Guillet}},\ and\ \bibinfo {author} {\bibfnamefont {G.}~\bibnamefont
  {Heinrich}},\ }\href {https://doi.org/10.1016/S0550-3213(00)00040-7}
  {\bibfield  {journal} {\bibinfo  {journal} {Nucl. Phys. B}\ }\textbf
  {\bibinfo {volume} {572}},\ \bibinfo {pages} {361} (\bibinfo {year}
  {2000})},\ \Eprint {https://arxiv.org/abs/hep-ph/9911342}
  {arXiv:hep-ph/9911342} \BibitemShut {NoStop}%
\bibitem [{\citenamefont {Chen}(2021)}]{Chen:2019wyb}%
  \BibitemOpen
  \bibfield  {author} {\bibinfo {author} {\bibfnamefont {L.}~\bibnamefont
  {Chen}},\ }\href {https://doi.org/10.1140/epjc/s10052-021-09210-9} {\bibfield
   {journal} {\bibinfo  {journal} {Eur. Phys. J. C}\ }\textbf {\bibinfo
  {volume} {81}},\ \bibinfo {pages} {417} (\bibinfo {year} {2021})},\ \Eprint
  {https://arxiv.org/abs/1904.00705} {arXiv:1904.00705 [hep-ph]} \BibitemShut
  {NoStop}%
\bibitem [{\citenamefont {Peraro}\ and\ \citenamefont
  {Tancredi}(2021)}]{Peraro:2020sfm}%
  \BibitemOpen
  \bibfield  {author} {\bibinfo {author} {\bibfnamefont {T.}~\bibnamefont
  {Peraro}}\ and\ \bibinfo {author} {\bibfnamefont {L.}~\bibnamefont
  {Tancredi}},\ }\href {https://doi.org/10.1103/PhysRevD.103.054042} {\bibfield
   {journal} {\bibinfo  {journal} {Phys. Rev. D}\ }\textbf {\bibinfo {volume}
  {103}},\ \bibinfo {pages} {054042} (\bibinfo {year} {2021})},\ \Eprint
  {https://arxiv.org/abs/2012.00820} {arXiv:2012.00820 [hep-ph]} \BibitemShut
  {NoStop}%
\bibitem [{\citenamefont {Lyubovitskij}\ \emph {et~al.}(2021)\citenamefont
  {Lyubovitskij}, \citenamefont {Wunder},\ and\ \citenamefont
  {Zhevlakov}}]{Lyubovitskij:2021ges}%
  \BibitemOpen
  \bibfield  {author} {\bibinfo {author} {\bibfnamefont {V.~E.}\ \bibnamefont
  {Lyubovitskij}}, \bibinfo {author} {\bibfnamefont {F.}~\bibnamefont
  {Wunder}},\ and\ \bibinfo {author} {\bibfnamefont {A.~S.}\ \bibnamefont
  {Zhevlakov}},\ }\href {https://doi.org/10.1007/JHEP06(2021)066} {\bibfield
  {journal} {\bibinfo  {journal} {JHEP}\ }\textbf {\bibinfo {volume} {06}},\
  \bibinfo {pages} {066}},\ \Eprint {https://arxiv.org/abs/2102.08943}
  {arXiv:2102.08943 [hep-ph]} \BibitemShut {NoStop}%
\bibitem [{\citenamefont {Anastasiou}\ \emph {et~al.}(2023)\citenamefont
  {Anastasiou}, \citenamefont {Karlen},\ and\ \citenamefont
  {Vicini}}]{Anastasiou:2023koq}%
  \BibitemOpen
  \bibfield  {author} {\bibinfo {author} {\bibfnamefont {C.}~\bibnamefont
  {Anastasiou}}, \bibinfo {author} {\bibfnamefont {J.}~\bibnamefont {Karlen}},\
  and\ \bibinfo {author} {\bibfnamefont {M.}~\bibnamefont {Vicini}},\ }\href
  {https://doi.org/10.1007/JHEP12(2023)169} {\bibfield  {journal} {\bibinfo
  {journal} {JHEP}\ }\textbf {\bibinfo {volume} {12}},\ \bibinfo {pages}
  {169}},\ \Eprint {https://arxiv.org/abs/2308.14701} {arXiv:2308.14701
  [hep-ph]} \BibitemShut {NoStop}%
\bibitem [{\citenamefont {Goode}\ \emph {et~al.}(2024)\citenamefont {Goode},
  \citenamefont {Herzog}, \citenamefont {Kennedy}, \citenamefont {Teale},\ and\
  \citenamefont {Vermaseren}}]{Goode:2024mci}%
  \BibitemOpen
  \bibfield  {author} {\bibinfo {author} {\bibfnamefont {J.}~\bibnamefont
  {Goode}}, \bibinfo {author} {\bibfnamefont {F.}~\bibnamefont {Herzog}},
  \bibinfo {author} {\bibfnamefont {A.}~\bibnamefont {Kennedy}}, \bibinfo
  {author} {\bibfnamefont {S.}~\bibnamefont {Teale}},\ and\ \bibinfo {author}
  {\bibfnamefont {J.}~\bibnamefont {Vermaseren}},\ }\href@noop {} {\  (\bibinfo
  {year} {2024})},\ \Eprint {https://arxiv.org/abs/2408.05137}
  {arXiv:2408.05137 [hep-ph]} \BibitemShut {NoStop}%
\bibitem [{\citenamefont {Chan}\ \emph {et~al.}(2012)\citenamefont {Chan},
  \citenamefont {Scoccimarro},\ and\ \citenamefont {Sheth}}]{Chan:2012jj}%
  \BibitemOpen
  \bibfield  {author} {\bibinfo {author} {\bibfnamefont {K.~C.}\ \bibnamefont
  {Chan}}, \bibinfo {author} {\bibfnamefont {R.}~\bibnamefont {Scoccimarro}},\
  and\ \bibinfo {author} {\bibfnamefont {R.~K.}\ \bibnamefont {Sheth}},\ }\href
  {https://doi.org/10.1103/PhysRevD.85.083509} {\bibfield  {journal} {\bibinfo
  {journal} {Phys. Rev. D}\ }\textbf {\bibinfo {volume} {85}},\ \bibinfo
  {pages} {083509} (\bibinfo {year} {2012})},\ \Eprint
  {https://arxiv.org/abs/1201.3614} {arXiv:1201.3614 [astro-ph.CO]}
  \BibitemShut {NoStop}%
\bibitem [{\citenamefont {Assassi}\ \emph {et~al.}(2014)\citenamefont
  {Assassi}, \citenamefont {Baumann}, \citenamefont {Green},\ and\
  \citenamefont {Zaldarriaga}}]{Assassi:2014fva}%
  \BibitemOpen
  \bibfield  {author} {\bibinfo {author} {\bibfnamefont {V.}~\bibnamefont
  {Assassi}}, \bibinfo {author} {\bibfnamefont {D.}~\bibnamefont {Baumann}},
  \bibinfo {author} {\bibfnamefont {D.}~\bibnamefont {Green}},\ and\ \bibinfo
  {author} {\bibfnamefont {M.}~\bibnamefont {Zaldarriaga}},\ }\href
  {https://doi.org/10.1088/1475-7516/2014/08/056} {\bibfield  {journal}
  {\bibinfo  {journal} {JCAP}\ }\textbf {\bibinfo {volume} {08}},\ \bibinfo
  {pages} {056}},\ \Eprint {https://arxiv.org/abs/1402.5916} {arXiv:1402.5916
  [astro-ph.CO]} \BibitemShut {NoStop}%
\bibitem [{\citenamefont {Senatore}(2015)}]{Senatore:2014eva}%
  \BibitemOpen
  \bibfield  {author} {\bibinfo {author} {\bibfnamefont {L.}~\bibnamefont
  {Senatore}},\ }\href {https://doi.org/10.1088/1475-7516/2015/11/007}
  {\bibfield  {journal} {\bibinfo  {journal} {JCAP}\ }\textbf {\bibinfo
  {volume} {11}},\ \bibinfo {pages} {007}},\ \Eprint
  {https://arxiv.org/abs/1406.7843} {arXiv:1406.7843 [astro-ph.CO]}
  \BibitemShut {NoStop}%
\bibitem [{\citenamefont {Desjacques}\ \emph
  {et~al.}(2018{\natexlab{a}})\citenamefont {Desjacques}, \citenamefont
  {Jeong},\ and\ \citenamefont {Schmidt}}]{Desjacques:2016bnm}%
  \BibitemOpen
  \bibfield  {author} {\bibinfo {author} {\bibfnamefont {V.}~\bibnamefont
  {Desjacques}}, \bibinfo {author} {\bibfnamefont {D.}~\bibnamefont {Jeong}},\
  and\ \bibinfo {author} {\bibfnamefont {F.}~\bibnamefont {Schmidt}},\ }\href
  {https://doi.org/10.1016/j.physrep.2017.12.002} {\bibfield  {journal}
  {\bibinfo  {journal} {Phys. Rept.}\ }\textbf {\bibinfo {volume} {733}},\
  \bibinfo {pages} {1} (\bibinfo {year} {2018}{\natexlab{a}})},\ \Eprint
  {https://arxiv.org/abs/1611.09787} {arXiv:1611.09787 [astro-ph.CO]}
  \BibitemShut {NoStop}%
\bibitem [{\citenamefont {Bernardeau}\ \emph {et~al.}(2002)\citenamefont
  {Bernardeau}, \citenamefont {Colombi}, \citenamefont {Gaztanaga},\ and\
  \citenamefont {Scoccimarro}}]{Bernardeau:2001qr}%
  \BibitemOpen
  \bibfield  {author} {\bibinfo {author} {\bibfnamefont {F.}~\bibnamefont
  {Bernardeau}}, \bibinfo {author} {\bibfnamefont {S.}~\bibnamefont {Colombi}},
  \bibinfo {author} {\bibfnamefont {E.}~\bibnamefont {Gaztanaga}},\ and\
  \bibinfo {author} {\bibfnamefont {R.}~\bibnamefont {Scoccimarro}},\ }\href
  {https://doi.org/10.1016/S0370-1573(02)00135-7} {\bibfield  {journal}
  {\bibinfo  {journal} {Phys. Rept.}\ }\textbf {\bibinfo {volume} {367}},\
  \bibinfo {pages} {1} (\bibinfo {year} {2002})},\ \Eprint
  {https://arxiv.org/abs/astro-ph/0112551} {arXiv:astro-ph/0112551}
  \BibitemShut {NoStop}%
\bibitem [{Note1()}]{Note1}%
  \BibitemOpen
  \bibinfo {note} {Assuming the Einstein-de Sitter approximation and neglecting
  scale-dependent effects due to massive neutrinos, the time and momentum
  dependencies factorize. The scale dependence can be incorporated using,
  e.g.,~the polynomial approximation of the growth factor~\cite
  {Levi:2016tlf,Chen:2021vba}.}\BibitemShut {Stop}%
\bibitem [{\citenamefont {Perko}\ \emph {et~al.}(2016)\citenamefont {Perko},
  \citenamefont {Senatore}, \citenamefont {Jennings},\ and\ \citenamefont
  {Wechsler}}]{Perko:2016puo}%
  \BibitemOpen
  \bibfield  {author} {\bibinfo {author} {\bibfnamefont {A.}~\bibnamefont
  {Perko}}, \bibinfo {author} {\bibfnamefont {L.}~\bibnamefont {Senatore}},
  \bibinfo {author} {\bibfnamefont {E.}~\bibnamefont {Jennings}},\ and\
  \bibinfo {author} {\bibfnamefont {R.~H.}\ \bibnamefont {Wechsler}},\
  }\href@noop {} {\  (\bibinfo {year} {2016})},\ \Eprint
  {https://arxiv.org/abs/1610.09321} {arXiv:1610.09321 [astro-ph.CO]}
  \BibitemShut {NoStop}%
\bibitem [{\citenamefont {Hamilton}(2000)}]{Hamilton:1999uv}%
  \BibitemOpen
  \bibfield  {author} {\bibinfo {author} {\bibfnamefont {A.~J.~S.}\
  \bibnamefont {Hamilton}},\ }\href
  {https://doi.org/10.1046/j.1365-8711.2000.03071.x} {\bibfield  {journal}
  {\bibinfo  {journal} {Mon. Not. Roy. Astron. Soc.}\ }\textbf {\bibinfo
  {volume} {312}},\ \bibinfo {pages} {257} (\bibinfo {year} {2000})},\ \Eprint
  {https://arxiv.org/abs/astro-ph/9905191} {arXiv:astro-ph/9905191}
  \BibitemShut {NoStop}%
\bibitem [{Note2()}]{Note2}%
  \BibitemOpen
  \bibinfo {note} {It is sufficient to consider numerators of the form
  $(\protect \hat {\protect \boldsymbol {q}}\cdot \protect \hat {\protect
  \boldsymbol {n}})^\ell $, since contraction with the external momenta can
  always be decomposed into pure scalar integrals using $\protect \hat
  {\protect \boldsymbol {q}}\cdot {\protect \boldsymbol {k}}_i = \protect \frac
  {1}{2q}(q^2+k_i^2-|{\protect \boldsymbol {q}}- {\protect \boldsymbol
  {k}}_i|^2)$.}\BibitemShut {Stop}%
\bibitem [{Note3()}]{Note3}%
  \BibitemOpen
  \bibinfo {note} {For standard galaxy biasing, multipoles up to $\ell =4$ and
  $\ell =6$ contribute to the one-loop power spectrum and bispectrum,
  respectively. However, additional selection effects, such as those for
  certain samples of galaxies~\cite {Desjacques:2018pfv, Tomlinson:2020xbf} and
  other tracers like the Lyman-alpha forest~\cite {Ivanov:2023yla}, necessitate
  including higher multipoles (up to $\ell =8$ for the one-loop power
  spectrum).}\BibitemShut {Stop}%
\bibitem [{\citenamefont {Funk}(1916)}]{Funk1916}%
  \BibitemOpen
  \bibfield  {author} {\bibinfo {author} {\bibfnamefont {P.}~\bibnamefont
  {Funk}},\ }\href {http://eudml.org/doc/158720} {\bibfield  {journal}
  {\bibinfo  {journal} {Mathematische Annalen}\ }\textbf {\bibinfo {volume}
  {77}},\ \bibinfo {pages} {136} (\bibinfo {year} {1916})}\BibitemShut
  {NoStop}%
\bibitem [{\citenamefont {Hecke}(1917)}]{Hecke1917}%
  \BibitemOpen
  \bibfield  {author} {\bibinfo {author} {\bibfnamefont {E.}~\bibnamefont
  {Hecke}},\ }\href {http://eudml.org/doc/158775} {\bibfield  {journal}
  {\bibinfo  {journal} {Mathematische Annalen}\ }\textbf {\bibinfo {volume}
  {78}},\ \bibinfo {pages} {398} (\bibinfo {year} {1917})}\BibitemShut
  {NoStop}%
\bibitem [{\citenamefont {Atkinson}\ and\ \citenamefont
  {Han}(2012)}]{atkinson2012spherical}%
  \BibitemOpen
  \bibfield  {author} {\bibinfo {author} {\bibfnamefont {K.}~\bibnamefont
  {Atkinson}}\ and\ \bibinfo {author} {\bibfnamefont {W.}~\bibnamefont {Han}},\
  }\href {https://books.google.com/books?id=SLQxcgt2Xj0C} {\emph {\bibinfo
  {title} {Spherical Harmonics and Approximations on the Unit Sphere: An
  Introduction}}},\ Lecture Notes in Mathematics\ (\bibinfo  {publisher}
  {Springer},\ \bibinfo {year} {2012})\BibitemShut {NoStop}%
\bibitem [{\citenamefont {Scoccimarro}(1997)}]{Scoccimarro:1996jy}%
  \BibitemOpen
  \bibfield  {author} {\bibinfo {author} {\bibfnamefont {R.}~\bibnamefont
  {Scoccimarro}},\ }\href {https://doi.org/10.1086/304578} {\bibfield
  {journal} {\bibinfo  {journal} {Astrophys. J.}\ }\textbf {\bibinfo {volume}
  {487}},\ \bibinfo {pages} {1} (\bibinfo {year} {1997})},\ \Eprint
  {https://arxiv.org/abs/astro-ph/9612207} {arXiv:astro-ph/9612207}
  \BibitemShut {NoStop}%
\bibitem [{\citenamefont {Linde}\ \emph {et~al.}(2024)\citenamefont {Linde},
  \citenamefont {Moradinezhad~Dizgah}, \citenamefont {Radermacher},
  \citenamefont {Casas},\ and\ \citenamefont {Lesgourgues}}]{Linde:2024uzr}%
  \BibitemOpen
  \bibfield  {author} {\bibinfo {author} {\bibfnamefont {D.}~\bibnamefont
  {Linde}}, \bibinfo {author} {\bibfnamefont {A.}~\bibnamefont
  {Moradinezhad~Dizgah}}, \bibinfo {author} {\bibfnamefont {C.}~\bibnamefont
  {Radermacher}}, \bibinfo {author} {\bibfnamefont {S.}~\bibnamefont {Casas}},\
  and\ \bibinfo {author} {\bibfnamefont {J.}~\bibnamefont {Lesgourgues}},\
  }\href@noop {} {\  (\bibinfo {year} {2024})},\ \Eprint
  {https://arxiv.org/abs/2402.09778} {arXiv:2402.09778 [astro-ph.CO]}
  \BibitemShut {NoStop}%
\bibitem [{Note4()}]{Note4}%
  \BibitemOpen
  \bibinfo {note} {To match these results requires using recursive relations
  between gamma functions. As a simple check, our ${\protect \sf I}_{\ell
  =1}^{\nu _1\nu _2}$ is related to $A_1=\protect \frac {1}{2}({\protect \sf
  I}_0^{\nu _1\nu _2}+{\protect \sf I}_0^{\nu _1-1,\nu _2}-{\protect \sf
  I}_0^{\nu _1,\nu _2-1})$ defined in \cite {Chudaykin:2020aoj} by ${\protect
  \sf I}_{1}^{\nu _1-\protect \frac {1}{2},\nu _2}=A_1$.}\BibitemShut {Stop}%
\bibitem [{\citenamefont {Varshalovich}\ \emph {et~al.}(1988)\citenamefont
  {Varshalovich}, \citenamefont {Moskalev},\ and\ \citenamefont
  {Khersonskii}}]{VMK}%
  \BibitemOpen
  \bibfield  {author} {\bibinfo {author} {\bibfnamefont {D.~A.}\ \bibnamefont
  {Varshalovich}}, \bibinfo {author} {\bibfnamefont {A.~N.}\ \bibnamefont
  {Moskalev}},\ and\ \bibinfo {author} {\bibfnamefont {V.~K.}\ \bibnamefont
  {Khersonskii}},\ }\href@noop {} {\emph {\bibinfo {title} {Quantum Theory of
  Angular Momentum}}}\ (\bibinfo  {publisher} {World Scientific},\ \bibinfo
  {address} {Singapore},\ \bibinfo {year} {1988})\BibitemShut {NoStop}%
\bibitem [{\citenamefont {Szapudi}(2004)}]{Szapudi:2004gh}%
  \BibitemOpen
  \bibfield  {author} {\bibinfo {author} {\bibfnamefont {I.}~\bibnamefont
  {Szapudi}},\ }\href {https://doi.org/10.1086/423168} {\bibfield  {journal}
  {\bibinfo  {journal} {Astrophys. J.}\ }\textbf {\bibinfo {volume} {614}},\
  \bibinfo {pages} {51} (\bibinfo {year} {2004})},\ \Eprint
  {https://arxiv.org/abs/astro-ph/0404477} {arXiv:astro-ph/0404477}
  \BibitemShut {NoStop}%
\bibitem [{Note5()}]{Note5}%
  \BibitemOpen
  \bibinfo {note} {This can also be written manifestly as a function of the
  three angles by expressing it as a sum over a product of three Legendre
  polynomials of these angles. However, it is computationally more efficient to
  fix a coordinate system and evaluate~\protect \textup {\hbox {\mathsurround
  \z@ \protect \normalfont (\ignorespaces \ref {TSH}\unskip \@@italiccorr
  )}}.}\BibitemShut {Stop}%
\bibitem [{Note6()}]{Note6}%
  \BibitemOpen
  \bibinfo {note} {We adopt a slightly different normalization compared
  to~\cite {VMK}. The tripolar scalar harmonic \protect \textup {\hbox
  {\mathsurround \z@ \protect \normalfont (\ignorespaces \ref {TSH}\unskip
  \@@italiccorr )}} consists of 3$j$ symbols and can be computed very
  efficiently using numerical algorithms such as the one presented in~\cite
  {Johansson:2015cca}.}\BibitemShut {Stop}%
\bibitem [{\citenamefont {Scoccimarro}\ \emph {et~al.}(1999)\citenamefont
  {Scoccimarro}, \citenamefont {Couchman},\ and\ \citenamefont
  {Frieman}}]{Scoccimarro:1999ed}%
  \BibitemOpen
  \bibfield  {author} {\bibinfo {author} {\bibfnamefont {R.}~\bibnamefont
  {Scoccimarro}}, \bibinfo {author} {\bibfnamefont {H.~M.~P.}\ \bibnamefont
  {Couchman}},\ and\ \bibinfo {author} {\bibfnamefont {J.~A.}\ \bibnamefont
  {Frieman}},\ }\href {https://doi.org/10.1086/307220} {\bibfield  {journal}
  {\bibinfo  {journal} {Astrophys. J.}\ }\textbf {\bibinfo {volume} {517}},\
  \bibinfo {pages} {531} (\bibinfo {year} {1999})},\ \Eprint
  {https://arxiv.org/abs/astro-ph/9808305} {arXiv:astro-ph/9808305}
  \BibitemShut {NoStop}%
\bibitem [{\citenamefont {Scoccimarro}(2015)}]{Scoccimarro:2015bla}%
  \BibitemOpen
  \bibfield  {author} {\bibinfo {author} {\bibfnamefont {R.}~\bibnamefont
  {Scoccimarro}},\ }\href {https://doi.org/10.1103/PhysRevD.92.083532}
  {\bibfield  {journal} {\bibinfo  {journal} {Phys. Rev. D}\ }\textbf {\bibinfo
  {volume} {92}},\ \bibinfo {pages} {083532} (\bibinfo {year} {2015})},\
  \Eprint {https://arxiv.org/abs/1506.02729} {arXiv:1506.02729 [astro-ph.CO]}
  \BibitemShut {NoStop}%
\bibitem [{\citenamefont {Yankelevich}\ and\ \citenamefont
  {Porciani}(2019)}]{Yankelevich:2018uaz}%
  \BibitemOpen
  \bibfield  {author} {\bibinfo {author} {\bibfnamefont {V.}~\bibnamefont
  {Yankelevich}}\ and\ \bibinfo {author} {\bibfnamefont {C.}~\bibnamefont
  {Porciani}},\ }\href {https://doi.org/10.1093/mnras/sty3143} {\bibfield
  {journal} {\bibinfo  {journal} {Mon. Not. Roy. Astron. Soc.}\ }\textbf
  {\bibinfo {volume} {483}},\ \bibinfo {pages} {2078} (\bibinfo {year}
  {2019})},\ \Eprint {https://arxiv.org/abs/1807.07076} {arXiv:1807.07076
  [astro-ph.CO]} \BibitemShut {NoStop}%
\bibitem [{\citenamefont {Agarwal}\ \emph {et~al.}(2021)\citenamefont
  {Agarwal}, \citenamefont {Desjacques}, \citenamefont {Jeong},\ and\
  \citenamefont {Schmidt}}]{Agarwal:2020lov}%
  \BibitemOpen
  \bibfield  {author} {\bibinfo {author} {\bibfnamefont {N.}~\bibnamefont
  {Agarwal}}, \bibinfo {author} {\bibfnamefont {V.}~\bibnamefont {Desjacques}},
  \bibinfo {author} {\bibfnamefont {D.}~\bibnamefont {Jeong}},\ and\ \bibinfo
  {author} {\bibfnamefont {F.}~\bibnamefont {Schmidt}},\ }\href
  {https://doi.org/10.1088/1475-7516/2021/03/021} {\bibfield  {journal}
  {\bibinfo  {journal} {JCAP}\ }\textbf {\bibinfo {volume} {03}},\ \bibinfo
  {pages} {021}},\ \Eprint {https://arxiv.org/abs/2007.04340} {arXiv:2007.04340
  [astro-ph.CO]} \BibitemShut {NoStop}%
\bibitem [{\citenamefont {Gualdi}\ and\ \citenamefont
  {Verde}(2020)}]{Gualdi:2020ymf}%
  \BibitemOpen
  \bibfield  {author} {\bibinfo {author} {\bibfnamefont {D.}~\bibnamefont
  {Gualdi}}\ and\ \bibinfo {author} {\bibfnamefont {L.}~\bibnamefont {Verde}},\
  }\href {https://doi.org/10.1088/1475-7516/2020/06/041} {\bibfield  {journal}
  {\bibinfo  {journal} {JCAP}\ }\textbf {\bibinfo {volume} {06}},\ \bibinfo
  {pages} {041}},\ \Eprint {https://arxiv.org/abs/2003.12075} {arXiv:2003.12075
  [astro-ph.CO]} \BibitemShut {NoStop}%
\bibitem [{\citenamefont {Pardede}\ \emph {et~al.}(2022)\citenamefont
  {Pardede}, \citenamefont {Rizzo}, \citenamefont {Biagetti}, \citenamefont
  {Castorina}, \citenamefont {Sefusatti},\ and\ \citenamefont
  {Monaco}}]{Pardede:2022udo}%
  \BibitemOpen
  \bibfield  {author} {\bibinfo {author} {\bibfnamefont {K.}~\bibnamefont
  {Pardede}}, \bibinfo {author} {\bibfnamefont {F.}~\bibnamefont {Rizzo}},
  \bibinfo {author} {\bibfnamefont {M.}~\bibnamefont {Biagetti}}, \bibinfo
  {author} {\bibfnamefont {E.}~\bibnamefont {Castorina}}, \bibinfo {author}
  {\bibfnamefont {E.}~\bibnamefont {Sefusatti}},\ and\ \bibinfo {author}
  {\bibfnamefont {P.}~\bibnamefont {Monaco}},\ }\href
  {https://doi.org/10.1088/1475-7516/2022/10/066} {\bibfield  {journal}
  {\bibinfo  {journal} {JCAP}\ }\textbf {\bibinfo {volume} {10}},\ \bibinfo
  {pages} {066}},\ \Eprint {https://arxiv.org/abs/2203.04174} {arXiv:2203.04174
  [astro-ph.CO]} \BibitemShut {NoStop}%
\bibitem [{\citenamefont {Amendola}\ \emph {et~al.}(2024)\citenamefont
  {Amendola}, \citenamefont {Marinucci}, \citenamefont {Pietroni},\ and\
  \citenamefont {Quartin}}]{Amendola:2023awr}%
  \BibitemOpen
  \bibfield  {author} {\bibinfo {author} {\bibfnamefont {L.}~\bibnamefont
  {Amendola}}, \bibinfo {author} {\bibfnamefont {M.}~\bibnamefont {Marinucci}},
  \bibinfo {author} {\bibfnamefont {M.}~\bibnamefont {Pietroni}},\ and\
  \bibinfo {author} {\bibfnamefont {M.}~\bibnamefont {Quartin}},\ }\href
  {https://doi.org/10.1088/1475-7516/2024/01/001} {\bibfield  {journal}
  {\bibinfo  {journal} {JCAP}\ }\textbf {\bibinfo {volume} {01}},\ \bibinfo
  {pages} {001}},\ \Eprint {https://arxiv.org/abs/2307.02117} {arXiv:2307.02117
  [astro-ph.CO]} \BibitemShut {NoStop}%
\bibitem [{\citenamefont {D'Amico}\ \emph {et~al.}(2024)\citenamefont
  {D'Amico}, \citenamefont {Donath}, \citenamefont {Lewandowski}, \citenamefont
  {Senatore},\ and\ \citenamefont {Zhang}}]{DAmico:2022osl}%
  \BibitemOpen
  \bibfield  {author} {\bibinfo {author} {\bibfnamefont {G.}~\bibnamefont
  {D'Amico}}, \bibinfo {author} {\bibfnamefont {Y.}~\bibnamefont {Donath}},
  \bibinfo {author} {\bibfnamefont {M.}~\bibnamefont {Lewandowski}}, \bibinfo
  {author} {\bibfnamefont {L.}~\bibnamefont {Senatore}},\ and\ \bibinfo
  {author} {\bibfnamefont {P.}~\bibnamefont {Zhang}},\ }\href
  {https://doi.org/10.1088/1475-7516/2024/05/059} {\bibfield  {journal}
  {\bibinfo  {journal} {JCAP}\ }\textbf {\bibinfo {volume} {05}},\ \bibinfo
  {pages} {059}},\ \Eprint {https://arxiv.org/abs/2206.08327} {arXiv:2206.08327
  [astro-ph.CO]} \BibitemShut {NoStop}%
\bibitem [{\citenamefont {Ivanov}\ \emph {et~al.}(2023)\citenamefont {Ivanov},
  \citenamefont {Philcox}, \citenamefont {Cabass}, \citenamefont {Nishimichi},
  \citenamefont {Simonovi\'c},\ and\ \citenamefont
  {Zaldarriaga}}]{Ivanov:2023qzb}%
  \BibitemOpen
  \bibfield  {author} {\bibinfo {author} {\bibfnamefont {M.~M.}\ \bibnamefont
  {Ivanov}}, \bibinfo {author} {\bibfnamefont {O.~H.~E.}\ \bibnamefont
  {Philcox}}, \bibinfo {author} {\bibfnamefont {G.}~\bibnamefont {Cabass}},
  \bibinfo {author} {\bibfnamefont {T.}~\bibnamefont {Nishimichi}}, \bibinfo
  {author} {\bibfnamefont {M.}~\bibnamefont {Simonovi\'c}},\ and\ \bibinfo
  {author} {\bibfnamefont {M.}~\bibnamefont {Zaldarriaga}},\ }\href
  {https://doi.org/10.1103/PhysRevD.107.083515} {\bibfield  {journal} {\bibinfo
   {journal} {Phys. Rev. D}\ }\textbf {\bibinfo {volume} {107}},\ \bibinfo
  {pages} {083515} (\bibinfo {year} {2023})},\ \Eprint
  {https://arxiv.org/abs/2302.04414} {arXiv:2302.04414 [astro-ph.CO]}
  \BibitemShut {NoStop}%
\bibitem [{\citenamefont {Chen}\ \emph {et~al.}(2024)\citenamefont {Chen},
  \citenamefont {Ivanov}, \citenamefont {Philcox},\ and\ \citenamefont
  {Wenzl}}]{Chen:2024vuf}%
  \BibitemOpen
  \bibfield  {author} {\bibinfo {author} {\bibfnamefont {S.-F.}\ \bibnamefont
  {Chen}}, \bibinfo {author} {\bibfnamefont {M.~M.}\ \bibnamefont {Ivanov}},
  \bibinfo {author} {\bibfnamefont {O.~H.~E.}\ \bibnamefont {Philcox}},\ and\
  \bibinfo {author} {\bibfnamefont {L.}~\bibnamefont {Wenzl}},\ }\href@noop {}
  {\  (\bibinfo {year} {2024})},\ \Eprint {https://arxiv.org/abs/2406.13388}
  {arXiv:2406.13388 [astro-ph.CO]} \BibitemShut {NoStop}%
\bibitem [{\citenamefont {Somogyi}(2011)}]{Somogyi:2011ir}%
  \BibitemOpen
  \bibfield  {author} {\bibinfo {author} {\bibfnamefont {G.}~\bibnamefont
  {Somogyi}},\ }\href {https://doi.org/10.1063/1.3615515} {\bibfield  {journal}
  {\bibinfo  {journal} {J. Math. Phys.}\ }\textbf {\bibinfo {volume} {52}},\
  \bibinfo {pages} {083501} (\bibinfo {year} {2011})},\ \Eprint
  {https://arxiv.org/abs/1101.3557} {arXiv:1101.3557 [hep-ph]} \BibitemShut
  {NoStop}%
\bibitem [{\citenamefont {Isono}\ \emph {et~al.}(2019)\citenamefont {Isono},
  \citenamefont {Noumi},\ and\ \citenamefont {Shiu}}]{Isono:2019wex}%
  \BibitemOpen
  \bibfield  {author} {\bibinfo {author} {\bibfnamefont {H.}~\bibnamefont
  {Isono}}, \bibinfo {author} {\bibfnamefont {T.}~\bibnamefont {Noumi}},\ and\
  \bibinfo {author} {\bibfnamefont {G.}~\bibnamefont {Shiu}},\ }\href
  {https://doi.org/10.1007/JHEP10(2019)183} {\bibfield  {journal} {\bibinfo
  {journal} {JHEP}\ }\textbf {\bibinfo {volume} {10}},\ \bibinfo {pages}
  {183}},\ \Eprint {https://arxiv.org/abs/1908.04572} {arXiv:1908.04572
  [hep-th]} \BibitemShut {NoStop}%
\bibitem [{\citenamefont {Levi}\ and\ \citenamefont
  {Vlah}(2016)}]{Levi:2016tlf}%
  \BibitemOpen
  \bibfield  {author} {\bibinfo {author} {\bibfnamefont {M.}~\bibnamefont
  {Levi}}\ and\ \bibinfo {author} {\bibfnamefont {Z.}~\bibnamefont {Vlah}},\
  }\href@noop {} {\  (\bibinfo {year} {2016})},\ \Eprint
  {https://arxiv.org/abs/1605.09417} {arXiv:1605.09417 [astro-ph.CO]}
  \BibitemShut {NoStop}%
\bibitem [{\citenamefont {Desjacques}\ \emph
  {et~al.}(2018{\natexlab{b}})\citenamefont {Desjacques}, \citenamefont
  {Jeong},\ and\ \citenamefont {Schmidt}}]{Desjacques:2018pfv}%
  \BibitemOpen
  \bibfield  {author} {\bibinfo {author} {\bibfnamefont {V.}~\bibnamefont
  {Desjacques}}, \bibinfo {author} {\bibfnamefont {D.}~\bibnamefont {Jeong}},\
  and\ \bibinfo {author} {\bibfnamefont {F.}~\bibnamefont {Schmidt}},\ }\href
  {https://doi.org/10.1088/1475-7516/2018/12/035} {\bibfield  {journal}
  {\bibinfo  {journal} {JCAP}\ }\textbf {\bibinfo {volume} {12}},\ \bibinfo
  {pages} {035}},\ \Eprint {https://arxiv.org/abs/1806.04015} {arXiv:1806.04015
  [astro-ph.CO]} \BibitemShut {NoStop}%
\bibitem [{\citenamefont {Tomlinson}\ \emph {et~al.}(2020)\citenamefont
  {Tomlinson}, \citenamefont {Gebhardt},\ and\ \citenamefont
  {Jeong}}]{Tomlinson:2020xbf}%
  \BibitemOpen
  \bibfield  {author} {\bibinfo {author} {\bibfnamefont {J.}~\bibnamefont
  {Tomlinson}}, \bibinfo {author} {\bibfnamefont {H.~S.~G.}\ \bibnamefont
  {Gebhardt}},\ and\ \bibinfo {author} {\bibfnamefont {D.}~\bibnamefont
  {Jeong}},\ }\href {https://doi.org/10.1103/PhysRevD.101.103528} {\bibfield
  {journal} {\bibinfo  {journal} {Phys. Rev. D}\ }\textbf {\bibinfo {volume}
  {101}},\ \bibinfo {pages} {103528} (\bibinfo {year} {2020})},\ \Eprint
  {https://arxiv.org/abs/2004.03629} {arXiv:2004.03629 [astro-ph.CO]}
  \BibitemShut {NoStop}%
\bibitem [{\citenamefont {Johansson}\ and\ \citenamefont
  {C.}(2016)}]{Johansson:2015cca}%
  \BibitemOpen
  \bibfield  {author} {\bibinfo {author} {\bibfnamefont {H.~T.}\ \bibnamefont
  {Johansson}}\ and\ \bibinfo {author} {\bibfnamefont {F.}~\bibnamefont {C.}},\
  }\href {https://doi.org/10.1137/15M1021908} {\bibfield  {journal} {\bibinfo
  {journal} {SIAM J. Sci. Comput.}\ }\textbf {\bibinfo {volume} {38}},\
  \bibinfo {pages} {A376} (\bibinfo {year} {2016})},\ \Eprint
  {https://arxiv.org/abs/1504.08329} {arXiv:1504.08329 [physics.comp-ph]}
  \BibitemShut {NoStop}%
\bibitem [{{\relax DLMF}()}]{NIST:DLMF}%
  \BibitemOpen
  {\relax DLMF},\ \href {https://dlmf.nist.gov/} {\bibinfo {title} {{\it NIST
  Digital Library of Mathematical Functions}}},\ \bibinfo {howpublished}
  {\url{https://dlmf.nist.gov/}, Release 1.2.1 of 2024-06-15},\ \bibinfo {note}
  {f.~W.~J. Olver, A.~B. {Olde Daalhuis}, D.~W. Lozier, B.~I. Schneider, R.~F.
  Boisvert, C.~W. Clark, B.~R. Miller, B.~V. Saunders, H.~S. Cohl, and M.~A.
  McClain, eds.}\BibitemShut {Stop}%
\end{thebibliography}%

\newpage

\appendix
\onecolumngrid

\section{One-Loop Master Integrals}\label{app:deriv}

In this appendix, we derive the main results of the paper: the master integrals for the one-loop power spectrum in \S III and the one-loop bispectrum in \S IV of the main text. 
Additionally, we present the formula for the one-loop trispectrum.

\subsection{One-Loop Power Spectrum}

The master integral for the one-loop power spectrum is
\begin{equation}
	I_\ell^{\nu_1\nu_2}(\k,\hat\n) =  \int_\q  \frac{\cL_\ell(\hat\q\cdot\hat\n)}{q^{2\nu_1}|\k-\q|^{2\nu_2}}=\int_0^\infty \frac{dq}{(2\pi)^3}\,q^{2-2\nu_1}\int_{S^2} d^2\hat\q\,   \frac{\cL_\ell(\hat\q\cdot\hat\n)}{|\k-\q|^{2\nu_2}}\,.\label{Iint2}
\end{equation}
We proceed by performing the radial and angular integrations separately. 
The angular integration can be easily done using the Funk-Heck formula~\eqref{FH} as
\begin{align}
	\int_{S^2} d^2\hat\q\, \frac{\cL_\ell(\hat\q \cdot\hat\n )}{|\hat\k-\q|^{2\nu_2}}
	&= \frac{4\pi}{2\ell+1}\sum_{m=-\ell}^\ell\sum_{m'=-\ell'}^{\ell'}\lambda_{\ell}^{\nu_2}(k,q)Y_{\ell'm'}^*(\hat\k) Y_{\ell m}(\hat\n)\int_{S^2} d^2\hat\q\,  Y_{\ell'm'}(\hat\q)Y_{\ell m}^*(\hat\q )\nn
	&= \frac{4\pi}{2\ell+1}\sum_{m=-\ell}^\ell\lambda_{\ell}^{\nu_2}(k,q)Y_{\ell m}^*(\hat\k) Y_{\ell m}(\hat\n)\nn
	&=\lambda_{\ell}^{\nu_2}(k,q)\cL_\ell(\hat\k\cdot\hat\n)\, ,
\end{align}  
where we used the orthonormality of the spherical harmonics and the addition theorem, and $\lambda_\ell^\nu$ was defined in \eqref{lam}. 
After rescaling $q\to q/k$, the remaining radial integral becomes dimensionless and can be explicitly evaluated as
\begin{align}
	\int_0^\infty \frac{d q}{(2\pi)^3}\,q^{2-2\nu_1}\lambda_{\ell}^{\nu_2}(1,q) &= \sum_{n=0}^\infty \frac{\Gamma(1-\nu_2)\Gamma(\frac{\ell+3}{2}+n-\nu_1)\Gamma(\frac{\ell-3}{2}+n+\nu_{12})}{(2\pi)^3n!\Gamma(\frac{3}{2}+\ell+n)}\nn
	&=\frac{1}{8\pi^{\frac{3}{2}}}\frac{\Gamma(\frac{3+\ell}{2}-\nu_1)\Gamma(\frac{3}{2}-\nu_2)\Gamma(\nu_{12}-\frac{3-\ell}{2})}{\Gamma(\nu_1+\frac{\ell}{2})\Gamma(\nu_2)\Gamma(3+\frac{\ell}{2}-\nu_{12})}\,,\label{Iintrad}
\end{align}
where we used the power series expansion of the hypergeometric function to perform the integral. 
Resumming the series then gives a product of gamma functions, reproducing the formula \eqref{sI} quoted in the main text.

It is useful to perform the integration in an alternative way, which will lead to a new identity for the hypergeometric function.
This identity will be important for deriving the formulas for the one-loop bispectrum and trispectrum, as shown in the subsequent sections.
Instead of directly integrating $\lambda_\ell^{\nu}$ over the entire range $[0,\infty)$, we split the radial integral as $\int_0^\infty = \int_0^1+\int_1^\infty$. 
In this case, the coefficient function $\lambda_\ell^{\nu}$ can be simplified in each integration region. 
Recall that
\begin{align}
	\lambda_{\ell}^\nu (k,q)
	&= \frac{2\pi^{\frac{3}{2}}\Gamma(1-\nu)}{\Gamma(\ell+\frac{3}{2})\Gamma(1-\nu-\ell)}\frac{(kq)^{\ell}}{(k^2+q^2)^{\ell+\nu}} \Gauss{\tfrac{\ell+\nu}{2}}{\tfrac{\ell+\nu+1}{2}}{\ell+\tfrac{3}{2}}{\frac{4k^2q^2}{(k^2+q^2)^2}}\,.\label{lamA}
\end{align}
Interestingly, this particular hypergeometric function is not completely generic and satisfies the following quadratic transformation formula~\cite[\href{https://dlmf.nist.gov/15.8\#E15}{(15.8.15)}]{NIST:DLMF}:
\begin{align}
	 \frac{1}{(1+z)^a}\,\Gauss{\frac{a}{2}}{\frac{a+1}{2}}{a-b+1}{\displaystyle\frac{4z}{(1+z)^2}}=\Gauss{a}{b}{a-b+1}{z}\,,\label{hyperid}
\end{align}
valid for $|z|<1$ and for $|z|=1$ when $\Re[b]< \frac{1}{2}$, where $a=\ell+\nu$ and $b=\nu-\frac{1}{2}$. 
Using this identity, we can express the coefficient function in a piecewise manner as
\begin{align}
	\lambda_{\ell}^\nu(k,q)	&=\begin{cases}
\, \widetilde\lambda_{\ell}^\nu(q,k) & (q\le  k)\\
\, \widetilde\lambda_{\ell}^\nu(k,q) & (k\le q)
 \end{cases}\,,
\end{align}
with
\begin{align}
	 \widetilde\lambda_{\ell}^\nu(k,q)&\equiv \frac{2\pi^{\frac{3}{2}}(1-\nu-\ell)_\ell}{\Gamma(\ell+\frac{3}{2})}k^{-2\nu}\left(\frac{q}{k}\right)^\ell\,\Gauss{\nu+\ell}{\nu-\frac{1}{2}}{\ell+\tfrac{3}{2}}{\frac{k^2}{q^2}}\,.\label{tlam}
\end{align}
This hypergeometric function also converges for $k=q$ when $\Re[\nu]<1$. 
Importantly, $\widetilde\lambda_{\ell}^\nu(q,k)$ and $\widetilde\lambda_{\ell}^\nu(k,q)$ are not analytic continuations of each other. 
Being a simple power series in $q$, the representation \eqref{tlam} is easier to integrate analytically than \eqref{lamA}.

As in standard dimensional regularization, we only keep the finite part of the integral. 
The contributions from $q=0,\infty$ then drop out, and we get
\begin{align}
		\sI_\ell (\nu_1,\nu_2) = A_1+A_2\,,
\end{align}
where
\begin{align}
	A_1 &\equiv \int_0^1 \frac{d q}{(2\pi)^3}\,q^{2-2\nu_1}\widetilde\lambda_{\ell}(\nu_2;1,q)=\frac{(-1)^\ell}{8\pi^{3/2}}\frac{\Gamma(c)\Gamma(\frac{1}{2}-b)}{\Gamma(1-a)}\HYPqR{a}{b}{c}{1+a-b}{1+c}{1}\,, \\
	A_2 &\equiv \int_1^\infty \frac{d q}{(2\pi)^3}\,q^{2-2\nu_1}\widetilde\lambda_{\ell}(\nu_2;q,1)=A_1|_{c\to a-c}\,,
\end{align}
with constants
\begin{align}
	a=\ell+\nu_2\,,\quad b=\nu_2-\frac{1}{2}\,,\quad c=\frac{3+\ell}{2}-\nu_1\,.
\end{align}
 This expresses $\sI_\ell$ as a linear combination of two ${}_3F_2$ functions, even though the final formula \eqref{Iintrad} is a simple ratio of a product of gamma functions.
In order for the two expressions to agree, the following identity must hold:
\begin{align}
	&\Gamma(c)\,\HYPqR{a}{b}{c}{1+a-b}{1+c}{1} +(c\to a-c)  = \frac{(-1)^{a-b}\cos(\pi b)\Gamma(1-a)\Gamma(1-b)\Gamma(c)\Gamma(a-c)}{i\pi\Gamma(1+a-b-c)\Gamma(1-b+c)}\,,\label{F3id}
\end{align}
for $\ell\in\mathbb{Z}$. 
Although not entirely apparent, it can be checked numerically that this identity holds.

\subsection{One-Loop Bispectrum}

The master integral for the one-loop bispectrum is
\begin{equation}
	J_\ell^{\nu_1\nu_2\nu_3}(\k_1,\k_2,\hat\n)\equiv\!\int_\q \frac{\cL_\ell(\hat\q\cdot\hat \n)}{q^{2\nu_1}|\k_1-\q|^{2\nu_2}|\k_2+\q|^{2\nu_3}}=\int_0^\infty \frac{dq}{(2\pi)^3}\,q^{2-2\nu_1}\int_{S^2} d^2\hat\q\,   \frac{\cL_\ell(\hat\q\cdot\hat\n)}{|\k-\q|^{2\nu_2}|\k_2+\q|^{2\nu_3}}\,.
\end{equation}
To evaluate this integral, we first project the individual angle-dependent factors onto spherical harmonics using the Funk-Hecke formula as
\begin{align}
	\cL_\ell(\hat\q\cdot\hat\n)&=\frac{4\pi}{2\ell+1}\sum_{m=-\ell}^\ell Y_{\ell m}^*(\hat\n)Y_{\ell m}(\hat\q)\,,\\
	|\k_1-\q|^{-2\nu_2} &= \sum_{L_2=0}^\infty \lambda_{L_2}^{\nu_2}(k_1,q) \sum_{M_2=-L_2}^{L_2} Y_{L_2M_2}^*(\hat\k_1)Y_{L_2M_2}(\hat\q)\,,\\
	|\k_2+\q|^{-2\nu_3} &= \sum_{L_3=0}^\infty (-1)^{L_3}\lambda_{L_3}^{\nu_3}(k_2,q) \sum_{M_3=-L_3}^{L_3} Y_{L_3M_3}^*(\hat\k_2)Y_{L_3M_3}(\hat\q)\,.
\end{align}
The angular integration of the three spherical harmonics over $\hat\q$ is then straightforward, which gives
\begin{align}
	\intS d^2\hat \q\, \frac{\cL_\ell(\hat\q\cdot\hat\n)}{|\k_1-\q|^{2\nu_2}|\k_2+\q|^{2\nu_3}}
	&= \sum_{L_2,L_3=0}^\infty(-1)^{L_3}\lambda_{L_2}^{\nu_2}(k_1,q)\lambda_{L_3}^{\nu_3}(k_2,q)S^{L_2L_3\ell}(\hat\k_1,\hat\k_2,\hat\n)\,,\nonumber
\end{align}
where $S^{L_2L_3\ell}$ is the tripolar spherical harmonics defined in \eqref{TSH}. The remaining dimensionless radial integral is
\begin{align}
	\sJ_{L_2L_3}(\nu_1,\nu_2,\nu_3; x)=(-1)^{L_3}\int_0^\infty \frac{dq}{(2\pi)^3}\, q^{2-2\nu_1}\lambda_{L_2}^{\nu_{2}}(1,q)\lambda_{L_3}^{\nu_{3}}(x,q)\,,
\end{align}
where $x\equiv k_2/k_1$. 
As before, we can perform the integral in a piecewise manner. 
To do so, instead of directly using $\lambda_\ell^\nu$, let us consider its piecewise representation $\tilde\lambda_\ell$ given in \eqref{tlam}. The integral can be split into three parts as $\int_0^\infty  = \int_0^x+\int_x^1+\int_1^\infty$ for $x\le 1$. The radial integral can then be expressed as
\begin{align}
	\sJ_{L_2L_3}(\nu_1,\nu_2,\nu_3; x) &= B_1(x;x)-B_2(x;x)+B_2(1;x)-B_3(1;x)\label{sJB}\,,
\end{align}
in terms of the indefinite integrals defined by
\begin{align}
	B_1(q;x) &=(-1)^{L_3}\int \frac{dq}{(2\pi)^3}\, q^{2-2\nu_1}\widetilde\lambda_{L_2}^{\nu_2}(1,q)\widetilde\lambda_{L_3}^{\nu_3}(x,q)\,,\\
	B_2(q;x) &=(-1)^{L_3}\int \frac{dq}{(2\pi)^3}\, q^{2-2\nu_1}\widetilde\lambda_{L_2}^{\nu_2}(1,q)\widetilde\lambda_{L_3}^{\nu_3}(q,x)\,,\\
	B_3(q;x) &=(-1)^{L_3}\int \frac{dq}{(2\pi)^3}\, q^{2-2\nu_1}\widetilde\lambda_{L_2}^{\nu_2}(q,1)\widetilde\lambda_{L_3}^{\nu_3}(q,x)\,.
\end{align}
We can then series expand one of the two $\widetilde\lambda$'s in each integral and perform the integration. For instance, the first two terms in \eqref{sJB} combine to give
\begin{align}
	B_1(x;x)-B_2(x;x) &=(-1)^{L_2}x^{3-2\nu_{13}+L_2}\frac{(1-L_2-\nu_2)_{L_2}(1-L_3-\nu_3)_{L_3}}{4}\label{B12diff1}\\
	&\times\sum_{n=0}^\infty \frac{(\nu_2-\frac{1}{2})_n (L_2+\nu_2)_n }{\Gamma(\frac{3}{2}+L_2+n)}\frac{x^{2n}}{n!}\left(\Gamma(c)\HYPqR{a}{b}{c}{1+a-b}{1+c}{1} + (c\to a-c) \right),\nonumber
\end{align}
where $a=L_3+\nu_3$, $b=\nu-\frac{1}{2}$, and $c=\frac{3+L_2+L_3}{2}-\nu_1+n$. Remarkably, this sum over ${}_3\widetilde F_2$ can be done analytically using the identity \eqref{F3id}. We get
\begin{align}
	B_1(x;x)-B_2(x;x) &=x^{3-2\nu_{13}}\frac{(-1)^{L_2+L_3}\sin(\pi\nu_3)}{4\cos(\pi(\nu_{13}+\frac{L_3-L_2}{2}))}\frac{\Gamma(1-\nu_1)\Gamma(1-\nu_2)\Gamma(\frac{3}{2}-\nu_2)\Gamma(\frac{3}{2}-\nu_3+\frac{L_2+L_3}{2})}{\Gamma(1-L_2-\nu_2)\Gamma(\nu_3+\frac{L_3-L_2}{2})}\nn
	&\quad\times x^{L_2}\HYPwR{\nu_2-\frac{1}{2}}{\nu_1+L_2}{1-\nu_3+\frac{L_2-L_3}{2}}{\frac{3}{2}-\nu_3+\frac{L_2+L_3}{2}}{\frac{3}{2}+L_2}{\frac{5}{2}-\nu_{13}+\frac{L_2-L_3}{2}}{3-\nu_{13}+\frac{L_2+L_3}{2}}{x^2},\label{B12diff2}
\end{align}
which gives the first term of the expression \eqref{sJ} shown in the main text. 
Similarly, it can be shown that the difference $B_2(1;x)-B_3(1;x)$ gives the second term of \eqref{sJ}.

\subsection{One-Loop Trispectrum}\label{app:tri}

The above calculation can be extended to  the one-loop trispectrum in redshift space. 
For previous numerical studies of the one-loop trispectrum in the EFT of LSS, see~\cite{Bertolini:2015fya, Bertolini:2016bmt, Steele:2021lnz}.
All of the one-loop trispectrum integrals can be computed from the master integral of the form
\begin{align}
	T^{\nu_1\nu_2\nu_3\nu_4}_\ell(\k_1,\k_2,\k_3,\hat\n) &= \int_\q \frac{\cL_\ell(\hat\q\cdot\hat\n)}{q^{2\nu_1}|\q-\k_1|^{2\nu_2}|\q+\k_2|^{2\nu_3}|\q+\k_{23}|^{2\nu_4}}\nn[2pt]
	&=\int_0^\infty \frac{dq}{(2\pi)^3}\,q^{2-2\nu_1}\int_{S^2}d^2\hat\q\, \frac{\cL_\ell(\hat\q\cdot\hat\n)}{|\q-\k_1|^{2\nu_2}|\q+\k_2|^{2\nu_3}|\q+\k_{23}|^{2\nu_4}}\,.
\end{align}
As before, we project the individual factors onto spherical harmonics to perform the angular integration. This gives
\begin{align}
	\cL_{\ell}(\hat\q\cdot\hat\n) &= \frac{4\pi}{2\ell+1}\sum_{m=-\ell}^\ell Y_{\ell m}(\hat\q)Y_{\ell m}^*(\hat\n)\,,\\
	|\q+\k_1|^{-2\nu_2} &= \sum_{L_2=0}^\infty (-1)^{L_2}\lambda_{L_2}^{\nu_2}(k,q) \sum_{M_2=-L_2}^{L_2}Y_{L_2M_2}(\hat\q)Y_{L_2M_2}^*(\hat\k_1)\,,\\
	|\q-\k_2|^{-2\nu_3} &= \sum_{L_3=0}^\infty \lambda_{L_3}^{\nu_3}(k_2,q) \sum_{M_3=-L_3}^{L_3}Y_{L_3M_3}(\hat\q)Y_{L_3M_3}^*(\hat\k_2)\,,\\
	|\q-\k_{23}|^{-2\nu_4} &= \sum_{L_4=0}^\infty \lambda_{L_4}^{\nu_4}(|\k_{23}|,q) \sum_{M_4=-L_4}^{L_4}Y_{L_4M_4}(\hat\q)Y_{L_4M_4}^*(\hat\k_{23})\,.
\end{align}
The angular integral then gives
\begin{align}
	&\int_{S^2}  d^2\hat\q\, Y_{\ell m}(\hat\q)Y_{L_2M_2}(\hat\q)Y_{L_3M_3}(\hat\q)Y_{L_4M_4}(\hat\q)= \sum_{L=0}^\infty \sum_{M=-L}^L (-1)^M \cG^{\ell L_2L}_{mM_2M}\cG^{L_3L_4L}_{M_3M_4(-M)}\,,
\end{align}
where we used
\begin{align}
	Y_{\ell_1 m_1}(\hat\q)Y_{\ell_2 m_2}(\hat\q) &= \sum_{L=0}^\infty \sum_{M=-L}^L \cG^{\ell_1\ell_2L}_{m_1m_2M} Y_{LM}^*(\hat\q)\,,\\
	\cG^{\ell_1\ell_2\ell_3}_{m_1m_2m_3}&=\sqrt{\frac{(2\ell_1+1)(2\ell_2+1)(2\ell_3+1)}{4\pi}}\begin{pmatrix}
		\ell_1 & \ell_2 & \ell_3 \\ 0 &0&0
	\end{pmatrix}\begin{pmatrix}
		\ell_1 & \ell_2 & \ell_3 \\ m_1 &m_2&m_3
	\end{pmatrix},\label{gaunt}
\end{align}
and $Y_{\ell m}^*(\hat\q) = (-1)^m Y_{\ell (-m)}(\hat\q)$. 
The full angular dependence is then captured by the function
\begin{align}
	&R^{L_2L_3L_4\ell}(\hat\k_1,\hat\k_2,\hat\k_{23},\hat\n)  \\
	&\ \equiv \frac{4\pi}{2\ell+1}\sum_{L=0}^\infty\sum_{M=-L}^L\sum_{M_2=-L_2}^{L_2}\sum_{M_3=-L_3}^{L_3}\sum_{M_4=-L_4}^{L_4}\sum_{m=-\ell}^\ell(-1)^M \cG^{\ell L_2L}_{mM_2M}\cG^{L_3L_4L}_{M_3M_4(-M)}Y_{\ell m}^*(\hat\n)Y_{L_2M_2}^*(\hat\k_1)Y_{L_3M_4}^*(\hat\k_2)Y_{L_4M_4}^*(\hat\k_{23})\,.\nonumber
\end{align}
Similar to \eqref{TSH}, this defines a {\it quadrupolar spherical harmonic},  which is a fully rotationally invariant, symmetric function of angles between four unit vectors. 
Factoring out an overall scaling in $k_1$, the master integral then becomes
\begin{align}
	T^{\nu_1\nu_2\nu_3\nu_4}_\ell(\k_1,\k_2,\k_3,\hat\n) &= k_1^{3-\nu_{1234}}\sum_{L_2,L_3,L_4=0}^\infty R^{L_2L_3L_4\ell}(\hat\k_1,\hat\k_2,\hat\k_{23},\hat\n) \sT_{L_2L_3L_4}^{\nu_1\nu_2\nu_3\nu_4} (x,y)\,,
\end{align}
where we have defined two ratios $x\equiv k_2/k_1$ and $y\equiv |\k_{23}|/k_1$, and the dimensionless radial integral is given by
\begin{align}
	\sT_{L_2L_3L_4}^{\nu_1\nu_2\nu_3\nu_4} (x,y) &\equiv (-1)^{L_2}\int_0^\infty \frac{dq}{(2\pi)^3}\, q^{2-2\nu_1} \lambda_{L_2}^{\nu_2}(1,q)\lambda_{L_3}^{\nu_3}(x,q)\lambda_{L_4}^{\nu_4}(y,q)\,.
\end{align}
Similar to the bispectrum case, this integral can be performed piecewisely. Without loss of generality, we can choose $x\le y\le 1$ and split the integral as $\int_0^\infty = \int_0^x + \int_x^y+\int_y^1+\int_1^\infty$. 
Going through similar (but slightly lengthier) steps as the bispectrum calculation, we find that the radial integral can be expressed as
\begin{align}
	\sT_{L_2L_3L_4}^{\nu_1\nu_2\nu_3\nu_4}(x,y)&=(-1)^{L_2}\Big(x^{3+L_{34}-2\nu_{12}}y^{-L_3-2\nu_3}{\mathsf t}_{L_2L_3L_4}^{(1)\nu_1\nu_2\nu_3\nu_4}(x,y)+x^{L_2}y^{3-L_2+L_4-2\nu_{123}}{\mathsf t}_{L_2L_3L_4}^{(2)\nu_1\nu_2\nu_3\nu_4}(x,y)\nn
	&\quad+x^{L_2}y^{L_3}{\mathsf t}_{L_2L_3L_4}^{(3)\nu_1\nu_2\nu_3\nu_4}(x,y)\Big)\,,
\end{align}
where 
\begin{align}
	{\mathsf t}_{L_2L_3L_4}^{(1)\nu_1\nu_2\nu_3\nu_4}(x,y) &= b_{234}\sum_{m=0}^\infty\frac{\Gamma(\frac{3+L_{234}}{2}-\nu_1+m)\Gamma(\nu_4-\frac{1}{2}+m)\Gamma(\nu_4+L_4+m)}{\cos(\frac{\pi(L_{34}-L_2+2(m-\nu_{12}))}{2})\Gamma(\frac{3}{2}+L_4+m)\Gamma(\frac{L_2-L_{34}}{2}+\nu_1-m)}\frac{x^{2m}}{m!}\nn
	&\hskip -35pt\times\Gamma(\nu_3-\tfrac{1}{2})\Gamma(\nu_3+L_3)\,\HYPwR{\nu_3-\frac{1}{2}}{\nu_3+L_3}{\frac{2-L_2+L_{34}+2m-2\nu_1}{2}}{\frac{3+L_{234}+2m-2\nu_1}{2}}{\frac{3}{2}+L_3}{\frac{5-L_2+L_{34}+2m-2\nu_{12}}{2}}{\frac{6+L_{234}+2m-2\nu_{12}}{2}}{\frac{x^2}{y^2}}\,,\\[5pt]
	{\mathsf t}_{L_2L_3L_4}^{(2)\nu_1\nu_2\nu_3\nu_4}(x,y) &=b_{342}\sum_{m=0}^\infty \frac{\Gamma(\frac{3-L_2+L_{34}}{2}-\nu_{12}-m)\Gamma(\nu_2-\frac{1}{2}+m)\Gamma(\nu_2+L_2+m)}{\cos(\frac{\pi(L_{23}-L_4+2(m+\nu_{123}))}{2})\Gamma(\frac{3}{2}+L_2+m)\Gamma(\frac{L_{23}-L_4}{2}+\nu_{12}+m)}\frac{(x/y)^{2m}}{m!}\nn
	&\hskip -35pt\times\Gamma(\nu_4-\tfrac{1}{2})\Gamma(\nu_4+L_4)\,\HYPwR{\nu_4-\frac{1}{2}}{\nu_4+L_4}{\frac{2-L_{23}+L_{4}-2m-2\nu_{12}}{2}}{\frac{3-L_2+L_{34}-2m-2\nu_{12}}{2}}{\frac{3}{2}+L_4}{\frac{5-L_{23}+L_4-2m-2\nu_{123}}{2}}{\frac{6-L_2+L_{34}-2m-2\nu_{123}}{2}}{y^2}\,,\\[5pt]
	{\mathsf t}_{L_2L_3L_4}^{(3)\nu_1\nu_2\nu_3\nu_4}(x,y) &= -b_{423}\sum_{m=0}^\infty \frac{\Gamma(\frac{-3+L_{234}}{2}+\nu_{1234}+m)\Gamma(\nu_3-\frac{1}{2}+m)\Gamma(\nu_3+L_3+m)}{\cos(\frac{\pi(L_{234}+2(m+\nu_{123}))}{2})\Gamma(\frac{3}{2}+L_3+m)\Gamma(\frac{-L_{23}+L_4}{2}+3-\nu_{1234}-m)}\frac{y^{2m}}{m!}\nn
	&\hskip -35pt\times\Gamma(\nu_2-\tfrac{1}{2})\Gamma(\nu_2+L_2)\,\HYPwR{\nu_2-\frac{1}{2}}{\nu_2+L_2}{\frac{-4+L_{23}-L_{4}+2m+2\nu_{1234}}{2}}{\frac{-3+L_{234}+2m+2\nu_{1234}}{2}}{\frac{3}{2}+L_2}{\frac{-1+L_{23}-L_4+2m+2\nu_{123}}{2}}{\frac{L_{234}+2m+2\nu_{123}}{2}}{x^2}\,,
\end{align}
with the prefactors given by
\begin{align}
	b_{i_1i_2i_3}\equiv \frac{4^{\nu_{i_1i_2i_3}}}{64\pi}\sin(\pi \nu_{i_1})\sin(2\pi\nu_{i_2})\sin(2\pi\nu_{i_3})\Gamma(2-2\nu_{i_1})\Gamma(2-2\nu_{i_2})\Gamma(2-2\nu_{i_3})\,.
\end{align}
Similar to the bispectrum case, for optimal efficiency, we can choose $x,y$ such that the ratios $x^2/y^2$, $x^2$, and $y^2$ remain small. 
For these configurations, the multipole sums converge rapidly, requiring only a few terms.

\end{document}